\documentclass[letterpaper]{article} 
\usepackage[preprint]{aaai2027}  
\usepackage[hyphens]{url}  
\usepackage{graphicx} 
\graphicspath{{./}{../}}
\usepackage{natbib}  
\usepackage{caption} 
\usepackage{algorithm}
\usepackage{algorithmic}
\usepackage{amsmath}
\usepackage{newfloat}
\usepackage{listings}
\DeclareCaptionStyle{ruled}{labelfont=normalfont,labelsep=colon,strut=off} 
\floatstyle{ruled}
\newfloat{listing}{tb}{lst}{}
\floatname{listing}{Listing}

\usepackage{inconsolata}          
\usepackage{upquote}              
\usepackage[skins,listings]{tcolorbox}
\tcbset{corpusbase/.style={%
  listing only, colback=black!5, colframe=black!45, arc=1.5mm, boxrule=0.4pt,
  left=2.5mm, right=2.5mm, top=1mm, bottom=1mm,
  listing options={basicstyle=\footnotesize\fontfamily{zi4}\selectfont,%
    numbers=none, breaklines=true, keepspaces=true, showstringspaces=false,%
    tabsize=2, literate={-}{{-}}1 {"}{{\textquotedbl}}1 {*}{{\raisebox{0.32ex}{*}}}1}}}
\newtcblisting{corpustree}{corpusbase}
\newtcblisting{corpusfile}[1]{corpusbase,%
  title={\ttfamily #1}, fonttitle=\footnotesize, colbacktitle=black!82,%
  coltitle=white, toptitle=0.4mm, bottomtitle=0.4mm}

\usepackage{booktabs}

\usepackage{array}

\definecolor{prjlink}{HTML}{001473}
\newcommand{\prj}[1]{\texttt{\textcolor{prjlink}{#1}}}

\usepackage{amssymb}
\usepackage{xcolor}
\usepackage{pgfplots}
\pgfplotsset{compat=1.18}

\copyrighttext{Disclaimer: Certain trade names and company products are mentioned in the text or identified. In no case does such identification i
mply recommendation or endorsement by the National Institute of Standards and Technology (NIST), nor that they are necessarily the best available
for the purpose.}

\title{CyberForge: Verified Vulnerability Injection at Repository Level for Cybersecurity Agent Training}

\author{
    Amine Lbath\textsuperscript{\rm 1,\rm 2}\equalcontrib,
    Manan Suri\textsuperscript{\rm 1,\rm 3}\equalcontrib,
    Aurelien Delaitre\textsuperscript{\rm 1},
    Vadim Okun\textsuperscript{\rm 1}, \\
    Massih-Reza Amini\textsuperscript{\rm 2},
    Ram D. Sriram\textsuperscript{\rm 1},
    Dinesh Manocha\textsuperscript{\rm 3}
}
\affiliations{
    \textsuperscript{\rm 1}National Institute of Standards and Technology,
    \textsuperscript{\rm 2}Université Grenoble Alpes, CNRS,
    \textsuperscript{\rm 3}University of Maryland, College Park
}

\begin{document}

\maketitle

\begin{abstract}

Despite recent advances, frontier large language model (LLM) agents remain limited in discovering and patching complex vulnerabilities in real-world software. Generally available agents can already aid attackers, who only need to find one exploitable weakness, while defenders must continuously identify and patch all vulnerabilities across fast-growing codebases. 
Stronger defensive agents would help close this gap, yet the scarcity of security training data with reproducible build and execution environments remains a bottleneck. 

We present CyberForge, a framework that synthesizes executable, repository-level security training data by injecting vulnerabilities into real C/C++ projects. It validates each instance dynamically: the injected build must pass the project's unit tests, and generated proof-of-vulnerability (PoV) must trigger on the injected build and not on the clean one.
CyberForge is not limited by the availability of disclosed vulnerabilities, therefore it can scale in comparison to data augmentation techniques which rely on historic CVE data. 
The resulting corpus holds 1\,034 validated vulnerabilities across 80 projects and 63 weakness categories, with edit locality similar to real CVE patches under a real-versus-real noise floor. 
Fine-tuning on trajectories collected over this corpus improves SEC-bench patch repair by
$+3.3$ to $+14.7$ points, in all six configurations of three model scales and two teachers, with the 31B student reaching its GPT-5.4-mini teacher, 72.7\,\% against 74.0\,\%. These gains generalize out of distribution to PatchEval, a corpus containing other programming languages, where every configuration also improves and the 31B student passes its teacher.

Code and data are available at \url{https://cyb3rforge.github.io}.
\end{abstract}

\section{Introduction}
\label{sec:intro}

Agentic systems have started finding real vulnerabilities. Anthropic's Mythos reported thousands of
zero-day findings across operating systems and browsers~\cite{mythos} and most recently, OpenAI disclosed that models under evaluation autonomously found and chained zero-days into a remote code execution path on Hugging Face's production infrastructure~\cite{openaihf}. On repository-level benchmarks that ask a model to locate a weakness in a real codebase and demonstrate it with an executable proof-of-vulnerability (PoV), even the strongest systems remain limited~\cite{mythos,exploitgym,secbench}.

The imbalance this creates favors attackers. An attacker needs one exploitable weakness and can retry indefinitely; a defender has to continuously secure an entire codebase. The balance shifts back only when automated detection and repair become dependable~\cite{frontierailandscape,teachaitohack}. Developing such capability in models that defenders can deploy and adapt is the goal of this work.

Availability of training data is the binding constraint~\cite{frontierailandscape,cyberzero}. Software-engineering agents benefited from datasets of packaged repositories with reproducible build and test environments, producing thousands of runnable instances and double-digit gains~\cite{swebench,swegym,swesmith,r2egym}. Security-agent training lacks a comparable pipeline because its validation problem is more complex. Functional bugs are exposed by tests: the buggy version fails at least one test that the fixed version passes. 
A vulnerability must instead remain \emph{latent} under existing tests and normal execution while still being \emph{triggerable} by adversarial input through a PoV that succeeds only on the vulnerable version. Automating this dual validation at scale, without a human expert in the loop, is what makes security data synthesis harder than its software-engineering analogue.

\begin{table*}[t]
\centering
\caption{Comparison of CyberForge with related work. Synthetic injection exists at function granularity (VGX, AVIATOR, ProSec) but targets detection models rather than agents and validates without execution. Existing project-level resources are evaluation artifacts whose growth is limited by the rate of public disclosure. No prior system supplies project-level \emph{training} data from synthetic injection validated by execution.}
\resizebox{0.8\textwidth}{!}{%
\begin{tabular}{llllll}
\toprule
\textbf{System} & \textbf{Purpose} & \textbf{Scope} & \textbf{Tasks} & \textbf{Bounded by disclosure} & \textbf{Validation} \\
\midrule
VGX~\cite{vgx}                 & Vuln.\ detection & Function & Detection              & $\times$   & Static \\
AVIATOR~\cite{aviator}         & Vuln.\ detection & Function & Detection              & $\times$   & Static \\
ProSec~\cite{prosec}           & Secure gen.      & Function & Alignment              & $\times$   & Static \\
\midrule
CTF-Dojo~\cite{ctfdojo}        & Training         & CTF      & Exploitation           & \checkmark & Execution \\
CyberZero~\cite{cyberzero}     & Training         & CTF      & Exploitation           & \checkmark & Simulation \\
\midrule
SEC-bench~\cite{secbench}      & Evaluation       & Project  & PoV, patch             & \checkmark & CVE replay \\
CyberGym~\cite{cybergym}       & Evaluation       & Project  & Exploitation           & \checkmark & CVE replay \\
BountyBench~\cite{bountybench} & Evaluation       & Project  & Detect, exploit, patch & \checkmark & Bug bounty \\
CVE-Genie~\cite{cvegenie}      & Evaluation       & Project  & Exploitation           & \checkmark & CVE replay \\
CVE-Factory~\cite{cvefactory}  & Training         & Project  & Patch                  & \checkmark & Automated CVE replay \\
\midrule
\textbf{CyberForge (ours)}     & \textbf{Training} & \textbf{Project} & \textbf{PoV, patch} & $\times$ & \textbf{Differential PoV} \\
\bottomrule
\end{tabular}%
}
\label{tab:comparison}
\end{table*}

Existing runnable vulnerability datasets are assembled from disclosed Common Vulnerabilities and Exposures (CVEs) and bug-bounty reports,
and are mostly built to evaluate rather than to train~\cite{secbench,cybergym,bountybench,cvegenie}. Each
disclosure has to be turned into an instance by reconstructing a historical environment, locating
the vulnerable state, and recovering a working PoV. Manual setup does not scale, automated replay
succeeds only in part~\cite{cybergym,secbench}. More fundamentally, these resources remain limited by the rate of human vulnerability discovery and public disclosure. Capture-the-flag tasks avoid that dependency but are built in idealized settings and transfer poorly to real software~\cite{cybergym,cvebench}. Function-level
injection techniques can be used to train LLMs to become vulnerability detectors, but transfer badly to realistic project level settings ~\cite{topscore, primevul, aviator}.

We introduce \textbf{CyberForge}, a scalable framework for generating executable, repository-level, PoV-validated security training data from real-world C/C++ projects. CyberForge creates new training instances by injecting candidate weaknesses into existing codebases. Rather than mining historical disclosures, this framework decouples corpus growth from the rate of human discovery and disclosure, while preserving the properties needed for agent training: realistic codebases, reproducible build and execution environments, and end-to-end PoV validation. 

CyberForge implements two complementary pipelines. \emph{Fuzzer-guided} injection uses OSS-Fuzz~\cite{ossfuzz} coverage to find sites reachable by an existing harness, so validation can lean on the harness itself. \emph{Agentic in-context injection} synthesizes weaknesses beyond harness reach. It combines a workflow of autonomous agents, static analysis tools, in-context examples, and iterative retry to synthesize and validate vulnerabilities end-to-end. Each instance operates on reproducible containerized environments derived from OSS-Fuzz images and is checked by a differential PoV oracle that requires the original and injected build to pass all tests and the PoV to trigger only on the injected build.

Across 80 C/C++ projects CyberForge produced 1\,034 validated vulnerabilities spanning 63
weakness categories. We fine-tune three open students, Gemma~4~31B, 12B and E4B~\cite{gemma4}, under two teachers, GPT-5.4-mini and Gemma~4~31B itself. All six configurations improve SEC-bench patch repair, by $+3.3$ to $+14.7$ points; the strongest raises the 31B student from 58.0\,\% to 72.7\,\%, comparable to the GPT-5.4-mini teacher that supervised it. The same training transferred to Go, JavaScript and Python repair on PatchEval, while the training corpus contains none of those languages. \textbf{Contributions:}
\begin{itemize}
\item \textbf{A repository-level generation framework.} CyberForge synthesizes cybersecurity-agent training data from buildable codebases through vulnerability injection and differential PoV validation, with two complementary pipelines.

\item \textbf{Validated corpus of project level vulnerabilties.} 1\,034 instances over 80 projects and 63 categories, with edit locality inside the noise floor separating two real CVE corpora.
\item \textbf{Downstream performance improvements.} SEC-bench score improves in all
six student--teacher configurations, by $+3.3$ to $+14.7$ points, and the gains hold on
PatchEval, an out-of-distribution cross-language benchmark.
\end{itemize}

\section{Related Work}
\label{sec:related}

Table~\ref{tab:comparison} compares CyberForge with prior vulnerability-generation, training and
evaluation systems.

SWE-bench~\cite{swebench} established repository-level evaluation from real pull requests, and SWE-Gym~\cite{swegym}, SWE-Smith~\cite{swesmith} and R2E-Gym~\cite{r2egym} turned the same idea into training data by synthesizing bugs inside runnable environments. SWE-Smith is the closest analogue to our injection pipeline, but its oracle does not apply to our case. It accepts an injection once a unit test fails, which signals that a functional bug was introduced. Our criterion is the opposite: the injected build must still pass every unit
test, and the weakness may surface only under a PoV input that leaves the clean build
unaffected.

Cyber-Zero~\cite{cyberzero} synthesizes agent trajectories without executable environments,
simulating runtime feedback from CTF writeups. CyberForge keeps execution in the loop, so every trajectory is grounded in a build that actually compiles and a PoV that actually triggers.

SEC-bench~\cite{secbench}, CyberGym~\cite{cybergym}, BountyBench~\cite{bountybench} and CVE-Genie~\cite{cvegenie} assemble runnable instances from disclosed vulnerabilities. They are evaluation artifacts, rather than scalable training pipelines, and remain limited by the supply of publicly disclosed vulnerabilities

VGX~\cite{vgx} and AVIATOR~\cite{aviator} inject weaknesses at function granularity to train detectors, validating statically or with learned filters rather than by execution. Neither produces a runnable project, and function-level supervision has transferred poorly to realistic detection~\cite{topscore,primevul}. CyberForge injects into repository-level buildable projects and dynamically validates instances. 

\section{CyberForge}
\label{sec:method}

\begin{figure*}[t]
  \centering
  \includegraphics[width=0.8\textwidth]{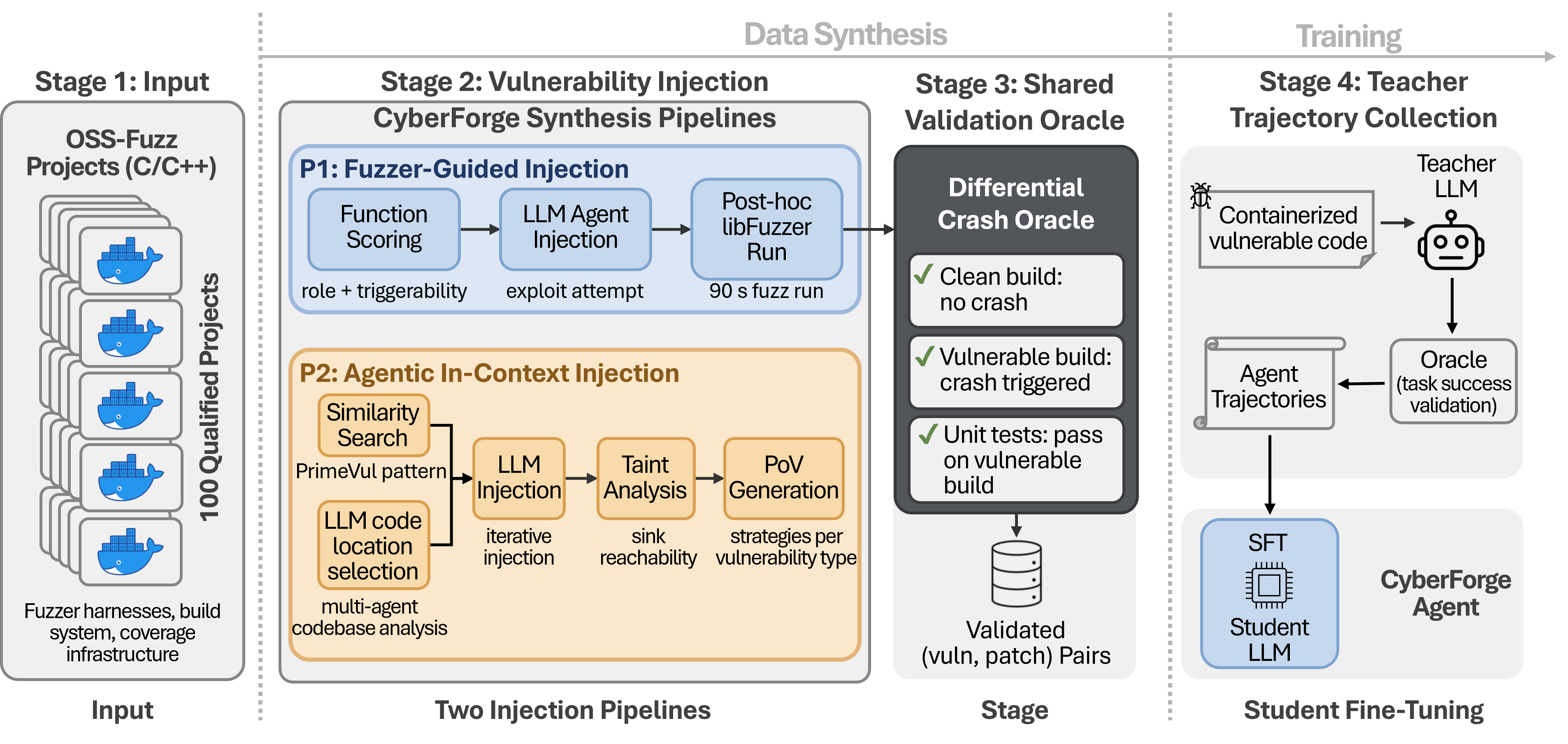}
  \caption{Overview of CyberForge. Two injection pipelines synthesize candidate vulnerable/patch
  pairs from OSS-Fuzz C/C++ projects; a shared differential PoV oracle admits a pair only if the
  injected build passes the project's tests and the PoV triggers on it alone; a teacher model collects verified trajectories for supervised fine-tuning.}
  \label{fig:overview}
\end{figure*}

Figure~\ref{fig:overview} presents CyberForge, an end-to-end vulnerability injection framework, that includes two complementary pipelines: 1) Fuzzer-guided Injection, which utilizes fuzzer metadata to prompt an LLM agent to inject the vulnerability, and 2) In-context Agentic Injection, which performs iterative injection in a specialized agent workflow. The generated vulnerabilities are validated using a differential oracle, after which the injections are used to mine teacher trajectories using capable models. 

\subsection{Project Setup}
\label{sec:setup}

\paragraph{Source and scope.}
CyberForge targets C/C++ projects enrolled in OSS-Fuzz~\cite{ossfuzz}. Two properties motivate the choice. Memory-safety defects, prevalent in these programming languages, remain the dominant class of critical CVEs in systems software~\cite{vuldet}, and OSS-Fuzz ships per-project Docker images with build scripts, sanitizer configuration and libFuzzer harnesses, which gives reproducible environments without per-project engineering.

\paragraph{Environment construction and qualification.}
For each candidate project, we build a container with a compiled binary, ASAN~\cite{asan} and UBSAN~\cite{ubsan} instrumentation, and the project's existing harnesses. No project-specific build logic is written, and every step is delegated to the project's preexisting \texttt{build.sh}. A project enters the pool only if its tests build and run unattended, pass at a 100\,\% rate on the unmodified codebase, with identical results across five runs. Projects with flaky tests are rejected. This process retained 100 qualified projects, 80 of which eventually contributed at least one validated instance.

\subsection{Vulnerability Injection}
\subsubsection{Pipeline 1: Fuzzer-Guided Injection}
\label{sec:p1}

This pipeline uses the existing fuzzer infrastructure of each OSS-Fuzz project to identify high-value injection sites, code locations that are both security-relevant and reachable by an existing libFuzzer harness, and then validates injections automatically, eliminating the need for a manually written PoV.

\paragraph{Step 1: Reachability extraction.}
For each project, we parse the offline OSS-Fuzz metadata cache, combining Fuzz Introspector reports, which summarize per-fuzzer reachability and coverage, with harness definitions and per-fuzzer coverage statistics to construct a reachability map of all functions reachable from at least one fuzzing harness.

\paragraph{Step 2: Candidate scoring and selection.}
Each reachable function is scored along two axes and ranked by a weighted combination of these scores. The first is a \emph{function score} based on reachability kind (runtime vs. static), structural role (e.g., parser entry point, buffer writer, decoder conversion, container access), call depth, fanout (the number of functions it directly calls), and file coverage. The second is a \emph{triggerability score} estimating how likely an injected fault at that site can be reached and triggered by an existing harness, using parser proximity, path signal strength, guard-to-sink distance (i.e., the proximity between a safety check and the operation it protects), and blocker proximity (i.e., the presence of nearby checks that may prevent malformed inputs from reaching the candidate site). Candidates are further diversified by capping per-file density and round-robin selecting across harness, role, and vulnerability category buckets to avoid duplication. Each selected candidate is annotated with an inferred weakness type (e.g., removed bounds check, unchecked index, copy-size constraint removal) and a confidence tier.

\paragraph{Step 3: LLM injection agent.}
The agent is given the candidate function, the precise site location and downstream operation within it, the inferred vulnerability category, the harness context, and the input format family (e.g., font binary, ZIP archive, UTF-16 text) derived from the harness subsystem. It also receives vulnerability-category-specific edit guidance specifying allowed modifications (e.g., operator widening, off-by-one bound change) and hard constraints (e.g., no new branches, single-file edit, preserve the downstream operation). The agent introduces one minimal, targeted modification that weakens the existing security check at the identified location. The modified project is compiled and run against unit tests; if either fails, the agent is prompted to retry.

\paragraph{Step 4: PoV generation and fuzzer-based validation.}
Once the injection passes unit tests, the agent produces a PoV script that exercises the injected fault. The PoV is guided by the input format context: the agent is provided with the input model, preferred seed extensions, and PoV-writing constraints derived from the harness. Given this context, the agent must demonstrate the vulnerability via a deterministic one-shot execution rather than open-ended fuzzing. This PoV is validated immediately by the shared differential PoV verifier (\S\ref{sec:validation}).

As a complementary validation path, a post-hoc libFuzzer run is also executed against the injected build. The run uses a format-aware seed corpus: format-specific input generators produce seeds matched to the harness's input model (e.g., GSUB table mutations for font parsers, ZIP header mutations for archive handlers), which are first replayed individually and then used to seed a timed libFuzzer run. Any input found by the fuzzer that triggers the vulnerable behavior on the new build is recorded as an additional confirmed PoV. Injections for which neither the agent PoV nor the fuzzer produces a trigger unique to the injected build are discarded.

\subsubsection{Pipeline 2: Agentic In-Context Injection}
\label{sec:p2}

This pipeline complements the other by synthesizing new vulnerabilities beyond those reachable by existing fuzz harness, drawing on candidate sites from two complementary selection strategies.

\paragraph{Candidate selection by hybrid retrieval.}
The first strategy matches the target project against the PrimeVul dataset of C/C++ functions
containing historical CVEs~\cite{primevul}, combining structural and semantic similarity retrieval ~\cite{hybridretrieval,codesearch}. A structural retriever compares functions by the $n$-grams of their AST node-type sequences~\cite{deckard}, so that matches reflect code topology rather than identifier names. A semantic retriever compares dense embeddings of the function text by cosine similarity. Their ranked lists are merged by reciprocal rank fusion~\cite{rrf}. A lightweight reranking step then demotes trivial accessor functions and isolated, uncalled helper functions. Each retained match supplies both an injection site in the target project and an aligned example pair (secure, vulnerable) that serves as the in-context template for the injection.

\paragraph{Candidate selection by agentic exploration.}
The second strategy lets specialist LLM agents explore the codebase directly, broadening diversity beyond sites that match known CVE patterns. A planner LLM splits a candidate budget across specialists, each responsible for one family of weaknesses, according to how relevant each family is to the project. Their proposals are then pooled, deduplicated, and ranked by an LLM score estimating how plausible, reachable, and project-relevant each proposed vulnerability is.

\paragraph{Injection stage.}
Each injection target is specified by the selected code location, the Common Weakness Enumeration (CWE) type, and its call chain and dataflow, recovered with the static-analysis tool CodeQL~\cite{codeql}. The agent modifies the project to introduce the vulnerability and produces an injection description. The modified project is then compiled and run against unit tests. Compilation or test failures trigger a revision loop back to this stage.

\paragraph{Taint analysis and PoV generation.}
The project is then analyzed using agent-based taint analysis to identify dataflow paths through which attacker-controlled input can reach the injected vulnerable location, providing valuable context on how to trigger the vulnerability. The PoV generation stage then constructs a proof-of-vulnerability using available signals: ASAN/UBSAN sanitizer reports, observable failure behavior, side effects (file writes, memory leaks), and output differences between the vulnerable and clean builds.

\paragraph{Verification and retry.}
The generated PoV is passed to a verifier, which runs the differential PoV oracle, and checks that the observed behavior is attributable to the injected vulnerability, with the sanitizer reporting the expected type of error at the expected code location, not an unrelated or spurious failure. If verification succeeds, the vulnerable code, its PoV, and the sanitizer report are saved. Otherwise, the loop retries PoV generation; after a fixed number of retries it loops back to the injection stage for a fresh injection.

\subsection{Differential Validation}
\label{sec:validation}

Both pipelines converge on a shared validation criterion.

\paragraph{Condition 1: Passing unit tests.}
A valid injection must not break any existing unit tests on the vulnerable build, as it must be \emph{latent}: present in the code but not triggered by normal execution paths, reflecting the nature of real-world security weaknesses that survive production testing and code review.

\paragraph{Condition 2: Differential PoV validation.}
The PoV must trigger on the injected build and not on the clean build under identical input. The condition validates injection and PoV jointly, since neither is meaningful without the other.

\subsection{Task Formation and Training}
\label{sec:task_formation}

Task construction matches SEC-bench~\cite{secbench}, following the same methodology as \cite{cvefactory}, so that training and evaluation align. The pipeline records full interaction traces from an agent running on this task in autonomy inside the generated dataset. Success is decided by the differential oracle of \S\ref{sec:validation} rather than by the agent's own report, so a trajectory that claims success without passing validation is discarded like any other failure. Applying the methodology from \cite{swegym, cvefactory}, only the successful trajectories are then used to train a model.

\section{Experimental Setup}
\label{sec:setup-exp}

\paragraph{Trajectory Collection}
For teacher trajectory extraction we used Mini-SWE-Agent \cite{minisweagent}, as a lightweight agentic scaffold. To prevent corrupting the run, the execution stays inside the project's OSS-Fuzz container with no network access, and no access to the reference patch, so it cannot recover the answer it is being trained to derive. 

\paragraph{Models and training.}
We fine-tune three open students, Gemma~4~31B, 12B and E4B~\cite{gemma4}, under two teachers: Gemma~4~31B itself, giving a self-distillation setting, and GPT-5.4-mini as a stronger teacher. Every run uses the same recipe and the same agentic scaffold at train and test time. We train LoRA adapters ($r{=}32$, $\alpha{=}64$, dropout $0.05$), at learning rate $1{\times}10^{-4}$ for three epochs on one H200.

\paragraph{Evaluation.}
The main benchmark is SEC-bench~\cite{secbench}, with 150 instances from 24 projects, evaluating patch success rates.
We also evaluated on PatchEval~\cite{patcheval}, as an out-of-distribution test, with 230 instances in Go, JavaScript and Python, none of which appear in our C/C++ corpus.

\section{Results}
\label{sec:results}

\subsection{Fine-Tuning on CyberForge Trajectories}
\label{sec:res-main}

Over the 1\,034 generated instances, we collected agent trajectories from two teachers: 1\,194 accepted trajectories from GPT-5.4-mini vs. 880 from Gemma~4~31B. Table~\ref{tab:combined-results} reports the performance of students trained on these trajectories. 

\textbf{Benchmark evaluation.}
On SEC-bench, all six student--teacher pairs gain from $+3.3$ to $+14.7$ points. The largest gain comes from Gemma~4~31B under a GPT-5.4-mini teacher, from 58.0\,\% to 72.7\,\%, within 1.3 points of the teacher. The Gemma~4~12B student roughly doubles from 8.7\,\% to 16.7\,\%.

\begin{table}[tb]
\centering
\caption{CyberForge, trained at different scales, with different teachers evaluated on SEC-bench (C/C++; in-domain), and PatchEval (Py, JS, Go; out-of-domain).}
\label{tab:combined-results}
\small
\setlength{\tabcolsep}{3.5pt}
\begin{tabular}{@{}lccc@{}}
\toprule
 & \textbf{SEC-bench} & \multicolumn{2}{c}{\textbf{PatchEval}} \\
\cmidrule(lr){3-4}
\textbf{Model} & (\%) & Strict (\%) & PoV (\%) \\
\midrule
\multicolumn{4}{@{}l}{\emph{Teachers (reference)}} \\
GPT-5.4-mini & 74.0 & 13.0 & 15.2 \\
Gemma 4 31B  & 58.0 & 12.2 & 14.4 \\
\midrule
Gemma 4 E4B (base) & 6.0 & 2.6 & 3.9 \\
\addlinespace[1pt]
\textbf{CyberForge-E4B} & \textbf{10.7}\,{\scriptsize$\uparrow$4.7} & 5.2\,{\scriptsize$\uparrow$2.6} & 6.5\,{\scriptsize$\uparrow$2.6} \\
\quad{\scriptsize\emph{Gemma 4 31B teacher}} \\
\textbf{CyberForge-E4B} & 9.3\,{\scriptsize$\uparrow$3.3} & \textbf{9.1}\,{\scriptsize$\uparrow$6.5} & \textbf{10.4}\,{\scriptsize$\uparrow$6.5} \\
\quad{\scriptsize\emph{GPT-5.4-mini teacher}} \\
\midrule
Gemma 4 12B (base) & 8.7 & 3.9 & 3.9 \\
\addlinespace[1pt]
\textbf{CyberForge-12B} & 16.0\,{\scriptsize$\uparrow$7.3} & 6.1\,{\scriptsize$\uparrow$2.2} & 8.7\,{\scriptsize$\uparrow$4.8} \\
\quad{\scriptsize\emph{Gemma 4 31B teacher}} \\
\textbf{CyberForge-12B} & \textbf{16.7}\,{\scriptsize$\uparrow$8.0} & \textbf{12.8}\,{\scriptsize$\uparrow$8.9} & \textbf{14.1}\,{\scriptsize$\uparrow$10.2} \\
\quad{\scriptsize\emph{GPT-5.4-mini teacher}} \\
\midrule
Gemma 4 31B (base) & 58.0 & 12.2 & 14.4 \\
\addlinespace[1pt]
\textbf{CyberForge-31B} & 64.7\,{\scriptsize$\uparrow$6.7} & 12.4\,{\scriptsize$\uparrow$0.2} & 15.7\,{\scriptsize$\uparrow$1.3} \\
\quad{\scriptsize\emph{Gemma 4 31B teacher}} \\
\textbf{CyberForge-31B} & \textbf{72.7}\,{\scriptsize$\uparrow$14.7} & \textbf{14.8}\,{\scriptsize$\uparrow$2.6} & \textbf{16.5}\,{\scriptsize$\uparrow$2.1} \\
\quad{\scriptsize\emph{GPT-5.4-mini teacher}} \\
\bottomrule
\end{tabular}
\end{table}

\begin{figure}[h]
\centering
\includegraphics[width=0.7\columnwidth]{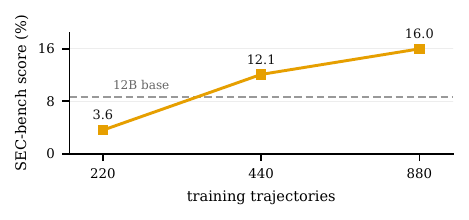}
\caption{ Training data scaling evaluated on SEC-bench.}
\label{fig:datascale}
\end{figure}

\textbf{Self-distillation and stronger teacher.}
On PatchEval the GPT-5.4-mini students beat the self-distilled ones at every scale, by 2.4 to 6.7 points. On SEC-bench the advantage grows with scale: 8.0 points at 31B, 0.7 at 12B, and at E4B it reverses, the Gemma teacher reaching 10.7\,\% against 9.3\,\%, both still beating their 6.0\,\% base. Notably, self-distillation helps at every scale, so the corpus carries signal that does not depend only on teacher's capabilities, paving the way for strong models learning from their own trajectories extracted through our workflow.

\begin{figure}[h]
\centering
\includegraphics[width=0.85\columnwidth]{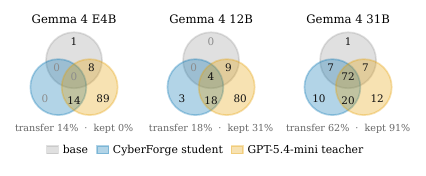}
\caption{Analysis of teacher transfer, and base retention on SEC-bench.}
\label{fig:venn}
\end{figure}

\textbf{Fine-tuning steers task resolution toward the teacher.}
Figure~\ref{fig:venn} splits each instance a student solves by where the ability came from: the student could already solve it before fine-tuning, its teacher can solve it, or neither can. Every student is steered in varying degree toward tasks its teacher can specifically solve. E4B is steered entirely onto them: all 14 of its solutions are ones the teacher also solves, and none of the nine its base solved survive, yet it picks up only 14\,\% of what the teacher can do and its base cannot. 31B gives up almost nothing, keeping 79 of its base's 87, and picks up 62\,\%, with 12B in between at 31\,\% kept and 18\,\% picked up. The smallest student is therefore steered the furthest and gains the least from it, which points to student capacity rather than the corpus as the limit. At 31B the student also moves past both sources, solving 10 instances that neither its base nor its teacher solves.

\textbf{The two teachers produce complementary students.}
Because the two students differ in which instances they solve, running both and keeping the successful run, recovers far more than either alone. Computed post hoc over the same runs and at the cost of a second inference pass, this ensemble scoring reaches 18.0, 25.3 and 82.0\,\% at E4B, 12B and 31B, 7.3 to 9.3 points above the better single student at every scale. The complementarity is significant enough to be worth more
than scale or supervision: two E4B students beat the best single 12B student, 18.0 against 16.7\,\%, and two 31B students beat the teacher that trained them, 82.0 against 74.0\,\%.

\textbf{The gains transfer out of distribution.}
PatchEval holds 230 CVEs in Go, JavaScript and Python, disjoint from the C/C++ training corpus, and scores each patch under a strict criterion and a PoV-blocking one. Every teacher--student pair improves on both: the 12B student by $+8.9$ strict and $+10.2$ PoV, and the 31B student reaches 14.8\,\% strict against the teacher's 13.0\,\%. This indicates that our training data enabled the model to generalize the concepts, rather than learning language-specific knowledge.

\textbf{Performance scales with corpus size.}
In Figure~\ref{fig:datascale}, the number of training trajectories varies, holding the student (Gemma~4~12B), the teacher (Gemma~4~31B) and the LoRA recipe fixed. Scores rise monotonically, from 3.6 to 12.1 to 16.0\,\% as the corpus doubles twice, with no sign of saturation at 880 trajectories. Importantly, at 220 trajectories the student scores \emph{below} its own base model, so a corpus too small to teach the workflow is worse than no fine-tuning at all.

\begin{table}[H]
\centering
\caption{Workflow ablation for Pipeline 2.}
\begin{tabular}{@{}lrr@{}}
\toprule
\textbf{Configuration} & \textbf{Injected} & \textbf{Validated} \\
\midrule
Naive single pass   & 68.2 & 0.0 \\
Taint analysis only & 75.5 & 2.8 \\
Retry loops only    & 77.6 & 3.5 \\
Full workflow       & 77.6 & \textbf{7.5} \\
\bottomrule
\end{tabular}
\label{tab:ablation}
\end{table}

\subsection{Generated Corpus}
\label{sec:res-corpus}

CyberForge made 16\,172 injection attempts, of which 1\,034 passed validation over 80 projects and 63 weakness categories. Writing a plausible injection is easy; producing one the differential oracle accepts is the hard part. Table~\ref{tab:ablation} ablates Pipeline 2: a naive single-pass agent already compiles and passes unit tests in 68.2\,\% of attempts, yet none of these pass validation, and the full workflow lifts the rate of plausible injections to 77.6\,\% while taking validated yield from 0\,\% to 7.5\,\%. Pipeline 1 shows the same from the validation side: a post-hoc fuzz-replay stage raises yield substantially with the injection stage untouched. The gain comes from the machinery around the injection, not from better injections.

\begin{figure}[h]
\centering
\includegraphics[width=0.8\columnwidth]{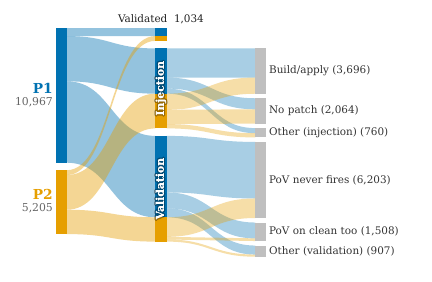}
\caption{Analysis of the Cyberforge data generation pipeline.}
\label{fig:sankey}
\end{figure}

Figure~\ref{fig:sankey} shows that the two pipelines fail at different stages: Pipeline 1 loses 64.2\,\% of its candidates at the validation stage and Pipeline 2 58.8\,\% at injection. A PoV that never triggers is the largest single cause of failure in both pipelines (44.6 and 33.1\,\%). Pipeline 2 additionally fails three times as often at producing a usable patch (24.2\,\% against 8.7\,\%), as its injection sites go beyond fuzzer reachability. 

We measured how closely the released instances resemble the real CVE patches in SEC-bench~\cite{secbench}, using the two-sample Kolmogorov--Smirnov distance~\cite{kstest} over the functions an edit touches. The distance is 0.165, against a 0.190 floor measured between two real CVE corpora, which supports the claim our injected vulnerabilities are near-realistic. The protocol is included in the supplement.

\begin{figure*}[tb]
\centering
\includegraphics[width=0.88\textwidth]{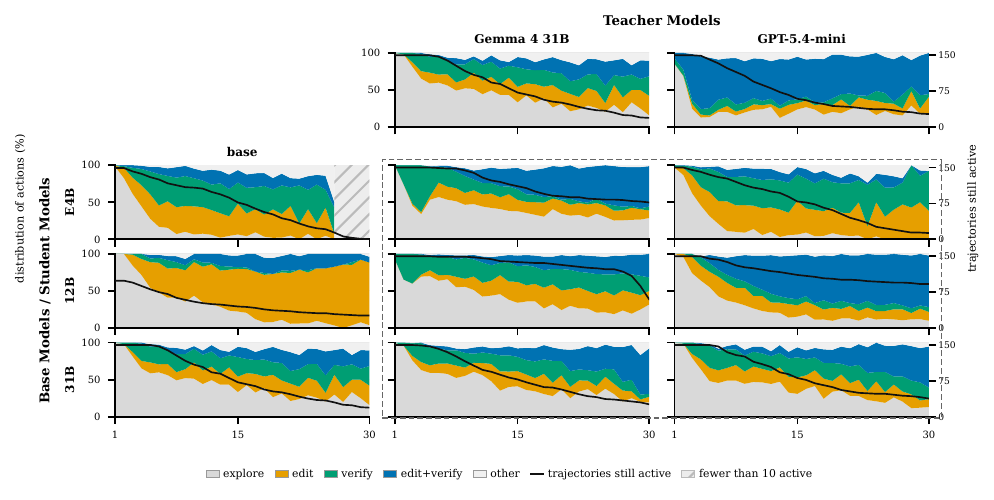}
\caption{ Analysis of coarse agent action behavior over 30 turns, demonstrating transfer of agent action density from teacher models to respective Cyberforge student models}
\label{fig:traj-temporal}
\end{figure*}

\subsection{How Fine-Tuning Changes Agent Behavior}
\label{sec:res-traj}

Each agent turn is one shell command, and a command can do several things at once, so a turn can take more than one label: \textsc{explore}, \textsc{edit}, \textsc{verify}, or the combination \textsc{edit+verify}, which we treat as its own state. Workflow coverage $c$ is the share of instances on which a \textsc{verify} occurs at or after an \textsc{edit}. Step budgets differ across runs, so every turn-indexed quantity uses the first 30 turns, the largest common horizon.

\begin{figure}[H]
\centering
\includegraphics[width=0.8\columnwidth]{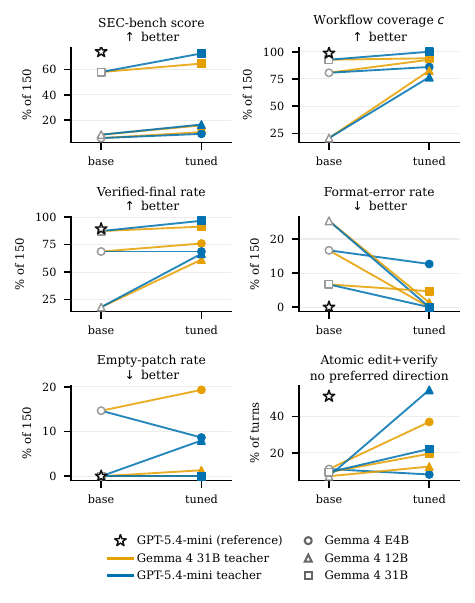}
\caption{Visualization of behavorial metrics comparing the base models (open markers) against fine-tuned CyberForge models (filled), with teacher models as anchors.}
\label{fig:traj-slope}
\end{figure}

\textbf{Students move toward their teacher's behavior.}
Figure~\ref{fig:traj-slope} compares each run's trajectory behavior between the base and fine-tuned model. Three of the four metrics measure the same thing: whether the agent finishes an edit\,$\rightarrow$\,verify cycle and returns something gradeable. Fine-tuning improves all three: each moves toward the teacher's value, the star, at every scale but one, E4B under the GPT teacher, which stays flat on the verified-final rate and still emits malformed output on 12.7\,\% of instances. The fourth, atomic \textsc{edit+verify}, tracks which teacher supervised the student, not its scale.

\textbf{Two independent channels.}
The SEC-bench score decomposes exactly as $\mathrm{score} = c\,q_v + (1-c)\,q_u$, with $q_v$ the solve
rate among instances that reached an edit\,$\rightarrow$\,verify cycle and $q_u$ the rate among the
rest. The two terms move independently and in opposite directions across scales. At 12B coverage
quadruples, $20.7 \rightarrow 82.7\,\%$, while $q_v$ is flat, so the gain comes entirely from the
model reaching the verification stage. At 31B coverage is already saturated and barely moves,
$92.7 \rightarrow 100.0\,\%$, while $q_v$ rises 10.1 points. The same training repairs whichever
channel was broken.

\paragraph{The 12B base edits without verifying.}
The mechanism behind the coverage gap is visible turn by turn
(Figure~\ref{fig:traj-temporal}). The 12B base model spends its trajectory issuing \textsc{edit}
commands, a median of 14 per trajectory among those that edit at all, while completing an
edit\,$\rightarrow$\,verify cycle on 20.7\,\% of instances with a median of \emph{zero} such cycles.
It edits blindly and never checks its work, and its active count falls fastest of any run. Both of
its fine-tunes acquire the verification band, and format violations collapse from 38 instances to 2
under the Gemma teacher and 0 under the GPT teacher.

\paragraph{Batching is a teacher fingerprint.}
GPT-5.4-mini writes and tests in a single command in 51\,\% of its turns, where base models do so in
7 to 11\,\% of theirs. Students distilled from it move toward it, 12B from 7 to 55\,\% and 31B from 9 to
22\,\%, while the Gemma-teacher students at the same scale do not. At 31B, the student taught by GPT
closes the gap with its teacher in behavior space as well as in score ($c = 100.0$ against 98.7\,\%,
$q_v = 72.7$ against 74.3\,\%): distillation transferred the workflow, not merely the accuracy.

\section{Conclusion}
\label{sec:conclusion}

We presented CyberForge, which builds security training data by injecting weaknesses into real C/C++ projects. Instances are verified through execution: the injected build must pass the project's own tests, and the proof of vulnerability must trigger on it but not on the clean build. Instances are created rather than mined, so corpus growth is decoupled from the rate of public disclosure. Across 80 OSS-Fuzz projects the two pipelines produced 1\,034 validated instances in 63 weakness categories, and the functions their edits touched are distributed about as locally as those of real vulnerabilities. 

Training on trajectories collected over this corpus improved SEC-bench in all six student--teacher configurations, by $+3.3$ to $+14.7$ points, with the 31B student reaching 72.7\,\% against its teacher's 74.0\,\%. The gains transfer across languages: every configuration also improved on PatchEval, which is entirely Go, JavaScript and Python while the training corpus is entirely C/C++. 

CyberForge gives defenders a scalable source of execution-validated data for training their own security agents.

\section*{Ethical Statement}

CyberForge aims to close the training-data gap on the \emph{defensive} side of AI-assisted security. All injected vulnerabilities are synthetic modifications of open-source code, not previously unknown bugs in deployed software. Models fine-tuned on CyberForge trajectories have improved cybersecurity capabilities. Our framework does not produce an exploit to actively operationalize a weakness and compromise a system. Instead, it generates a proof of vulnerability (PoV) that demonstrates a weakness exists and is reachable, without delivering a malicious payload. We note that similar capabilities already exist in other models; our contribution is to study the role of targeted training data, not to introduce capabilities that do not otherwise exist.

\bibliography{document}

\appendix

\section{The CyberForge Corpus}
\label{app:corpus}

This section describes the released corpus. An instance is one injected weakness together with the proof of vulnerability that triggers it, admitted only after the differential oracle accepted the pair. Table~\ref{tab:app-summary} provides an overview of our generated corpus.

\begin{table}[H]
\centering
\caption{The CyberForge corpus at a glance.}
\small
\begin{tabular}{@{}lrr@{}}
\toprule
\textbf{Corpus} & & \\
\midrule
Validated instances                   & \multicolumn{2}{r}{1\,034} \\
\quad Pipeline 1 (fuzzer-guided)      & \multicolumn{2}{r}{643} \\
\quad Pipeline 2 (agentic)            & \multicolumn{2}{r}{391} \\
OSS-Fuzz projects qualified           & \multicolumn{2}{r}{100} \\
\quad contributing $\geq 1$ instance  & \multicolumn{2}{r}{80} \\
Distinct weakness categories (CWE)    & \multicolumn{2}{r}{63} \\
Distinct CWE groups                   & \multicolumn{2}{r}{25} \\
\midrule
\textbf{By language} & \textbf{Projects} & \textbf{Instances} \\
\midrule
C++                                   & 73 & 697 \\
C                                     & 27 & 337 \\
\bottomrule
\end{tabular}
\label{tab:app-summary}
\end{table}

\subsection{Projects}

CyberForge targets C and C++ projects from OSS-Fuzz. A project enters the pool only if its own test suite builds, runs unattended and passes at a 100\,\% rate on unmodified code. This left 100 qualified projects, 80 of which contributed at least one validated instance. Table~\ref{tab:app-projects} lists them.

\begin{table*}[tp]
\centering
\caption{All 100 qualified OSS-Fuzz projects. \textbf{L}: primary language. \textbf{Type}: application domain. \textbf{Stars}: GitHub stars where applicable. \textbf{Files} and \textbf{KLoC}: C/C++ source files and thousands of lines in the project's own tree, measured over compiled sources. \textbf{P1}/\textbf{P2}: instances from the fuzzer-guided and agentic pipelines. \textbf{CWE}: distinct weakness categories present. \textbf{UT}: tests in the project's own suite.}
\scriptsize
\setlength{\tabcolsep}{2pt}
\begin{tabular}{@{}lllr@{\hspace{6pt}}rrrrrr@{}}
\toprule
\textbf{Project} & \textbf{L} & \textbf{Type} & \textbf{Stars} & \textbf{Files} & \textbf{KLoC} & \textbf{P1} & \textbf{P2} & \textbf{CWE} & \textbf{UT} \\
\midrule
\prj{arrow} & C++ & Data & 16\,968 & 6 & 4 & 0 & 9 & 6 & 11 \\
\prj{assimp} & C++ & Media & 13\,101 & 636 & 300 & 12 & 10 & 9 & 584 \\
\prj{behaviortreecpp} & C++ & Utility & 4\,135 & 134 & 373 & 0 & 3 & 3 & 13 \\
\prj{binutils} & C++ & System & -- & 1\,603 & 2\,013 & 0 & 0 & 0 & 222 \\
\prj{bluez} & C & Network & 1\,120 & 455 & 316 & 7 & 4 & 7 & 37 \\
\prj{brotli} & C++ & Compr. & 14\,820 & 111 & 43 & 29 & 10 & 11 & 73 \\
\prj{c-blosc} & C++ & Compr. & 1\,056 & 143 & 92 & 6 & 7 & 6 & 1\,643 \\
\prj{c-blosc2} & C++ & Compr. & 581 & 431 & 140 & 0 & 12 & 5 & 2\,017 \\
\prj{cgif} & C & Media & 150 & 73 & 7 & 7 & 5 & 6 & 60 \\
\prj{cjson} & C++ & Data & 12\,890 & 28 & 11 & 7 & 23 & 15 & 19 \\
\prj{coturn} & C & Network & 14\,261 & 51 & 26 & 5 & 16 & 13 & 16 \\
\prj{cryptofuzz} & C++ & Crypto & 26 & 228 & 135 & 0 & 0 & 0 & 5 \\
\prj{dnsmasq} & C & Network & -- & 11 & 10 & 0 & 6 & 5 & 48 \\
\prj{double-conversion} & C++ & Numeric & 1\,193 & 39 & 329 & 34 & 5 & 5 & 9 \\
\prj{envoy} & C++ & Network & 28\,669 & 893 & 184 & 0 & 0 & 0 & 8 \\
\prj{espeak-ng} & C++ & Text & 6\,696 & 112 & 57 & 8 & 8 & 7 & 19 \\
\prj{exiv2} & C++ & Media & 1\,145 & 201 & 107 & 3 & 3 & 5 & 6 \\
\prj{ffmpeg} & C++ & Media & 62\,591 & 1\,057 & 337 & 0 & 0 & 0 & 2\,779 \\
\prj{file} & C++ & Utility & 1\,639 & 37 & 22 & 1 & 0 & 1 & 86 \\
\prj{flac} & C++ & Media & 2\,371 & 156 & 70 & 4 & 4 & 6 & 26 \\
\prj{flex} & C & Runtime & 4\,036 & 35 & 31 & 0 & 3 & 3 & 203 \\
\prj{fluent-bit} & C++ & Network & 8\,001 & 2\,198 & 1\,344 & 6 & 1 & 3 & 63 \\
\prj{fmt} & C++ & Utility & 23\,696 & 68 & 70 & 29 & 11 & 12 & 21 \\
\prj{freerdp} & C & Network & 13\,508 & 1\,462 & 534 & 0 & 3 & 3 & 155 \\
\prj{ghostscript} & C++ & Doc & -- & 2\,078 & 2\,086 & 0 & 1 & 1 & 6 \\
\prj{grok} & C++ & Media & 290 & 967 & 437 & 0 & 7 & 4 & 177 \\
\prj{guetzli} & C++ & Media & 12\,917 & 48 & 10 & 33 & 13 & 8 & 10 \\
\prj{h2o} & C++ & Network & 11\,521 & 355 & 164 & 0 & 6 & 4 & 37 \\
\prj{h3} & C & Numeric & 6\,432 & 174 & 36 & 67 & 17 & 14 & 315 \\
\prj{haproxy} & C++ & Network & 6\,742 & 467 & 303 & 0 & 3 & 3 & 4 \\
\prj{harfbuzz} & C++ & Text & 5\,957 & 399 & 166 & 27 & 5 & 8 & 66 \\
\prj{hdf5} & C & Data & 962 & 1\,279 & 1\,140 & 5 & 0 & 2 & 2\,804 \\
\prj{htslib} & C++ & Data & 939 & 159 & 117 & 8 & 2 & 5 & 353 \\
\prj{hunspell} & C++ & Text & 2\,550 & 56 & 90 & 4 & 12 & 5 & 139 \\
\prj{icu} & C++ & Text & 3\,563 & 1\,134 & 589 & 6 & 3 & 6 & 21 \\
\prj{jq} & C & Data & 35\,308 & 57 & 39 & 20 & 13 & 11 & 9 \\
\prj{json-c} & C++ & Data & 3\,286 & 137 & 17 & 9 & 5 & 9 & 26 \\
\prj{jsoncpp} & C++ & Data & 8\,877 & 30 & 14 & 0 & 2 & 2 & 124 \\
\prj{jsonnet} & C++ & Runtime & 7\,549 & 104 & 124 & 0 & 2 & 2 & 59 \\
\prj{kamailio} & C & Network & 2\,891 & 420 & 185 & 20 & 3 & 8 & 45 \\
\prj{kimageformats} & C++ & Media & -- & 87 & 41 & 0 & 0 & 0 & 45 \\
\prj{lcms} & C++ & Media & 730 & 35 & 44 & 17 & 1 & 5 & 148 \\
\prj{leptonica} & C++ & Media & 2\,068 & 568 & 324 & 2 & 8 & 7 & 121 \\
\prj{libarchive} & C++ & Compr. & 3\,572 & 995 & 243 & 0 & 6 & 4 & 897 \\
\prj{libdwarf} & C & System & 256 & 284 & 148 & 0 & 3 & 3 & 22 \\
\prj{libjxl} & C++ & Media & 3\,610 & 590 & 196 & 10 & 9 & 7 & 13 \\
\prj{liblouis} & C & Text & 338 & 81 & 53 & 0 & 0 & 0 & 18 \\
\prj{libpng} & C++ & Media & 1\,636 & 36 & 42 & 22 & 2 & 6 & 36 \\
\prj{libredwg} & C & Data & 1\,513 & 446 & 990 & 0 & 0 & 0 & 254 \\
\prj{libsass} & C++ & Runtime & 4\,325 & 148 & 40 & 9 & 17 & 13 & 27 \\
\bottomrule
\end{tabular}
\hfill
\begin{tabular}{@{}lllr@{\hspace{6pt}}rrrrrr@{}}
\toprule
\textbf{Project} & \textbf{L} & \textbf{Type} & \textbf{Stars} & \textbf{Files} & \textbf{KLoC} & \textbf{P1} & \textbf{P2} & \textbf{CWE} & \textbf{UT} \\
\midrule
\prj{libsndfile} & C & Media & 1\,705 & 133 & 65 & 2 & 0 & 2 & 143 \\
\prj{libsrtp} & C++ & Crypto & 1\,394 & 69 & 30 & 14 & 2 & 6 & 11 \\
\prj{libucl} & C & Data & 1\,739 & 22 & 19 & 2 & 7 & 6 & 8 \\
\prj{libvips} & C++ & Media & 11\,540 & 464 & 258 & 0 & 2 & 2 & 7 \\
\prj{libxslt} & C++ & Doc & -- & 63 & 51 & 0 & 1 & 1 & 10 \\
\prj{libyang} & C & Data & 426 & 332 & 213 & 5 & 2 & 4 & 61 \\
\prj{libzip} & C++ & Compr. & 1\,033 & 217 & 23 & 18 & 3 & 9 & 183 \\
\prj{lighttpd} & C & Network & 702 & 169 & 100 & 0 & 2 & 2 & 232 \\
\prj{llvm} & C++ & Runtime & 39\,572 & 1\,948 & 805 & 0 & 0 & 0 & 10 \\
\prj{mbedtls} & C++ & Crypto & 6\,845 & 566 & 649 & 2 & 0 & 1 & 133 \\
\prj{mruby} & C++ & Runtime & 5\,597 & 266 & 180 & 3 & 6 & 7 & 1\,773 \\
\prj{njs} & C++ & Runtime & 1\,588 & 121 & 94 & 4 & 3 & 4 & 6\,024 \\
\prj{ntopng} & C++ & Network & 8\,047 & 124 & 57 & 0 & 0 & 0 & 17 \\
\prj{open5gs} & C & Network & 2\,657 & 7\,526 & 1\,475 & 10 & 1 & 5 & 3 \\
\prj{open62541} & C++ & Network & 3\,192 & 199 & 130 & 0 & 4 & 4 & 15 \\
\prj{opencv} & C++ & Media & 90\,237 & 88 & 44 & 5 & 2 & 4 & 135 \\
\prj{openh264} & C++ & Media & 6\,125 & 217 & 83 & 1 & 5 & 5 & 631 \\
\prj{openjph} & C++ & Media & 291 & 76 & 41 & 0 & 3 & 3 & 82 \\
\prj{opensc} & C++ & Crypto & 3\,061 & 301 & 216 & 0 & 2 & 2 & 5 \\
\prj{openssl} & C & Crypto & 30\,536 & 1\,702 & 606 & 9 & 3 & 6 & 26 \\
\prj{openthread} & C++ & Network & 3\,999 & 1\,089 & 560 & 0 & 3 & 3 & 9 \\
\prj{openvpn} & C & Network & 14\,321 & 181 & 93 & 0 & 0 & 0 & 17 \\
\prj{pcapplusplus} & C++ & Network & 3\,121 & 329 & 144 & 0 & 4 & 4 & 304 \\
\prj{pcre2} & C++ & Utility & 1\,329 & 62 & 109 & 12 & 0 & 4 & 4 \\
\prj{perfetto} & C++ & System & -- & -- & -- & 9 & 1 & 2 & 649 \\
\prj{php} & C++ & Runtime & 40\,267 & 1\,096 & 1\,377 & 0 & 0 & 0 & 816 \\
\prj{poppler} & C++ & Doc & -- & 373 & 194 & 0 & 0 & 0 & 15 \\
\prj{protobuf-c} & C & Data & 2\,985 & 39 & 71 & 17 & 1 & 6 & 11 \\
\prj{pupnp} & C & Network & 437 & 140 & 56 & 20 & 4 & 6 & 14 \\
\prj{qemu} & C & System & -- & 4\,007 & 1\,483 & 0 & 0 & 0 & 213 \\
\prj{qpid-proton} & C++ & Network & 249 & 105 & 37 & 0 & 0 & 0 & 29 \\
\prj{quickjs} & C & Runtime & 10\,889 & 23 & 80 & 28 & 13 & 16 & 8 \\
\prj{radare2} & C++ & System & 24\,462 & 1\,180 & 643 & 0 & 2 & 1 & 575 \\
\prj{relic} & C++ & Crypto & 513 & 386 & 166 & 0 & 1 & 1 & 19 \\
\prj{ruby} & C++ & Runtime & 23\,666 & 358 & 187 & 0 & 0 & 0 & 894 \\
\prj{selinux} & C & System & 1\,612 & 275 & 110 & 4 & 3 & 5 & 68 \\
\prj{serenity} & C++ & System & 33\,706 & 7 & 2 & 26 & 2 & 4 & 7 \\
\prj{simdjson} & C++ & Data & 24\,102 & 7 & 2 & 17 & 1 & 5 & 118 \\
\prj{skia} & C++ & Media & -- & 619 & 165 & 0 & 0 & 0 & 8 \\
\prj{systemd} & C++ & System & 16\,540 & 2\,452 & 827 & 7 & 3 & 6 & 15 \\
\prj{unbound} & C & Network & 4\,750 & 215 & 188 & 0 & 0 & 0 & 88 \\
\prj{unicorn} & C++ & System & 9\,201 & 516 & 451 & 4 & 0 & 2 & 11 \\
\prj{uwebsockets} & C++ & Network & 18\,937 & 57 & 16 & 0 & 1 & 1 & 7 \\
\prj{wamr} & C & Runtime & 6\,043 & 130 & 117 & 0 & 0 & 0 & 148 \\
\prj{wireshark} & C++ & Network & -- & 2\,966 & 5\,746 & 0 & 13 & 8 & 9 \\
\prj{wuffs} & C++ & Media & 4\,795 & 14 & 95 & 1 & 2 & 2 & 78 \\
\prj{xz} & C++ & Compr. & 1\,624 & 148 & 39 & 0 & 6 & 3 & 19 \\
\prj{yajl-ruby} & C++ & Data & 1\,489 & 15 & 3 & 0 & 0 & 0 & 416 \\
\prj{zeek} & C++ & Network & 7\,830 & 2\,542 & 1\,156 & 0 & 0 & 0 & 113 \\
\prj{zlib-ng} & C++ & Compr. & 2\,056 & 177 & 62 & 6 & 0 & 2 & 70 \\
\bottomrule
\end{tabular}
\label{tab:app-projects}
\end{table*}

\subsection{Weakness Coverage}
\begin{figure*}[t]
\centering
\includegraphics[width=0.9\textwidth]{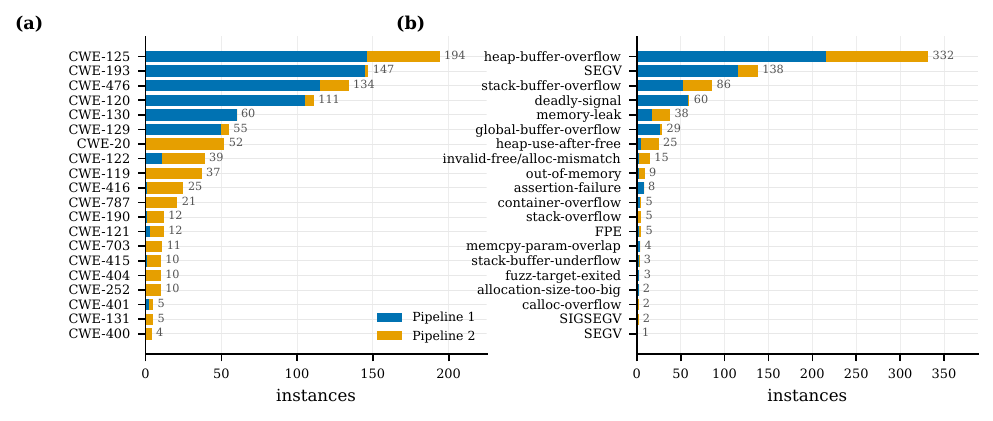}
\caption{What the corpus covers, by pipeline. (a) The 20 most frequent CWE
identifiers; Table~\ref{tab:app-cwes} gives their names. (b) The 20 most frequent
runtime fault classes in the sanitizer reports.}
\label{fig:app-coverage}
\end{figure*}

The corpus covers 63 CWE types across 25 classes. Table~\ref{tab:app-cwes} enumerates every CWE type and Figure~\ref{fig:app-coverage}(a) shows the most frequently injected. Memory-safety weaknesses dominate, as they do among real C/C++ disclosures.

The two injection pipelines cover different CWE types, so running both increases coverage and diversity. The fuzzer-guided pipeline covers 15 CWE types and relies on the paths recorded by existing libFuzzer harnesses, which favor memory safety weaknesses. The agentic pipeline covers 62 distinct CWE types. It selects sites using static analysis and retrieval, which allow for more freedom regarding the types of vulnerability it can inject. The pipelines' coverage overlap on 14 CWE types. 46 projects received instances from both pipelines, 7 from the fuzzer-guided pipeline alone, and 27 solely from the agentic pipeline.

\begin{table}[H]
\centering
\caption{Weakness categories realized in the corpus, grouped by the coarse family the injection pipeline selects over. Groups are ordered by instance count and identifiers within a group by their own instance count.}
\scriptsize
\setlength{\tabcolsep}{4pt}
\begin{tabular}{@{}l >{\raggedright\arraybackslash}p{5.5cm}@{}}
\toprule
\textbf{Group} & \textbf{CWE identifiers realized} \\
\midrule
Post buffer operation & CWE-125, CWE-120, CWE-130, CWE-129, CWE-122, CWE-119, CWE-787, CWE-121, CWE-170, CWE-126, CWE-680, CWE-823 \\
Calculation & CWE-193, CWE-131, CWE-369 \\
Invalid pointer & CWE-476, CWE-824 \\
Web & CWE-20 \\
Expired memory & CWE-416, CWE-415 \\
Numeric errors & CWE-190, CWE-191, CWE-189, CWE-197, CWE-682 \\
Resource management & CWE-404, CWE-400, CWE-399, CWE-666, CWE-770 \\
Unhandled errors & CWE-703, CWE-391, CWE-754 \\
Return value & CWE-252, CWE-690 \\
Input validation & CWE-78, CWE-74, CWE-77, CWE-95 \\
Memory leak & CWE-401 \\
Other & CWE-863, CWE-203, CWE-354, CWE-532 \\
Access control & CWE-264, CWE-284 \\
Confidentiality & CWE-200, CWE-209 \\
Path-related & CWE-22 \\
Concurrency & CWE-362 \\
Memory release & CWE-590, CWE-762 \\
Privileges & CWE-269, CWE-271 \\
Initialization & CWE-457 \\
Loop and recursion & CWE-674 \\
Strings & CWE-134 \\
API & CWE-475 \\
Control flow & CWE-670 \\
Encapsulation & CWE-485 \\
Function call & CWE-227 \\
(none recorded) & CWE-617, CWE-665 \\
\bottomrule
\end{tabular}
\label{tab:app-groups}
\end{table}

\begin{table*}[tp]
\centering
\caption{All 63 weakness categories realized in the corpus, ordered by instance count. Names follow MITRE. \textbf{P1}/\textbf{P2}: instances per pipeline, with the Pipeline 2 count split by selection strategy into \textbf{E} (agentic exploration) and \textbf{R} (retrieval). \textbf{Pr.}: projects in which the category occurs.}
\scriptsize
\setlength{\tabcolsep}{3pt}
\begin{tabular}{@{}l >{\raggedright\arraybackslash}p{11cm} rrrrrr@{}}
\toprule
\textbf{CWE} & \textbf{Name} & \textbf{All} & \textbf{P1} & \textbf{P2} &
\textbf{E} & \textbf{R} & \textbf{Pr.} \\
\midrule
CWE-125 & Out-of-bounds Read & 194 & 146 & 48 & 14 & 34 & 46 \\
CWE-193 & Off-by-one Error & 147 & 145 & 2 & 1 & 1 & 38 \\
CWE-476 & NULL Pointer Dereference & 134 & 115 & 19 & 7 & 12 & 41 \\
CWE-120 & Buffer Copy without Checking Size of Input ('Classic Buffer Overflow') & 111 & 105 & 6 & 3 & 3 & 35 \\
CWE-130 & Improper Handling of Length Parameter Inconsistency & 60 & 60 & 0 & 0 & 0 & 10 \\
CWE-129 & Improper Validation of Array Index & 55 & 50 & 5 & 5 & 0 & 22 \\
CWE-20 & Improper Input Validation & 52 & 0 & 52 & 48 & 4 & 24 \\
CWE-122 & Heap-based Buffer Overflow & 39 & 11 & 28 & 12 & 16 & 22 \\
CWE-119 & Improper Restriction of Operations within the Bounds of a Memory Buffer & 37 & 0 & 37 & 13 & 24 & 24 \\
CWE-416 & Use After Free & 25 & 1 & 24 & 6 & 18 & 18 \\
CWE-787 & Out-of-bounds Write & 21 & 0 & 21 & 3 & 18 & 16 \\
CWE-121 & Stack-based Buffer Overflow & 12 & 3 & 9 & 4 & 5 & 9 \\
CWE-190 & Integer Overflow or Wraparound & 12 & 1 & 11 & 4 & 7 & 9 \\
CWE-703 & Improper Check or Handling of Exceptional Conditions & 11 & 0 & 11 & 0 & 11 & 8 \\
CWE-252 & Unchecked Return Value & 10 & 0 & 10 & 10 & 0 & 6 \\
CWE-404 & Improper Resource Shutdown or Release & 10 & 0 & 10 & 10 & 0 & 2 \\
CWE-415 & Double Free & 10 & 1 & 9 & 5 & 4 & 7 \\
CWE-131 & Incorrect Calculation of Buffer Size & 5 & 0 & 5 & 5 & 0 & 4 \\
CWE-401 & Improper Release of Memory Before Removing Last Reference ('Memory Leak') & 5 & 2 & 3 & 1 & 2 & 5 \\
CWE-191 & Integer Underflow (Wrap or Wraparound) & 4 & 0 & 4 & 1 & 3 & 3 \\
CWE-22 & Improper Limitation of a Pathname to a Restricted Directory ('Path Traversal') & 4 & 0 & 4 & 2 & 2 & 4 \\
CWE-400 & Uncontrolled Resource Consumption ('Resource Exhaustion') & 4 & 0 & 4 & 3 & 1 & 3 \\
CWE-690 & Unchecked Return Value to NULL Pointer Dereference & 4 & 0 & 4 & 4 & 0 & 4 \\
CWE-189 & Numeric Errors & 3 & 0 & 3 & 0 & 3 & 3 \\
CWE-200 & Information Exposure & 3 & 0 & 3 & 1 & 2 & 3 \\
CWE-362 & Concurrent Execution using Shared Resource with Improper Synchronization & 3 & 0 & 3 & 0 & 3 & 2 \\
CWE-369 & Divide By Zero & 3 & 1 & 2 & 1 & 1 & 3 \\
CWE-399 & Resource Management Errors & 3 & 0 & 3 & 0 & 3 & 3 \\
CWE-78 & Improper Neutralization of Special Elements used in an OS Command  & 3 & 0 & 3 & 3 & 0 & 3 \\
CWE-134 & Uncontrolled Format String & 2 & 0 & 2 & 0 & 2 & 2 \\
CWE-170 & Improper Null Termination & 2 & 0 & 2 & 2 & 0 & 2 \\
CWE-197 & Numeric Truncation Error & 2 & 0 & 2 & 2 & 0 & 2 \\
CWE-264 & Permissions, Privileges, and Access Controls & 2 & 0 & 2 & 0 & 2 & 1 \\
CWE-269 & Improper Privilege Management & 2 & 0 & 2 & 0 & 2 & 2 \\
CWE-284 & Improper Access Control & 2 & 0 & 2 & 2 & 0 & 2 \\
CWE-391 & Unchecked Error Condition & 2 & 0 & 2 & 2 & 0 & 1 \\
CWE-457 & Use of Uninitialized Variable & 2 & 0 & 2 & 1 & 1 & 1 \\
CWE-590 & Free of Memory not on the Heap & 2 & 1 & 1 & 1 & 0 & 2 \\
CWE-617 & Reachable Assertion & 2 & 0 & 2 & 2 & 0 & 2 \\
CWE-666 & Operation on Resource in Wrong Phase of Lifetime & 2 & 0 & 2 & 2 & 0 & 1 \\
CWE-674 & Uncontrolled Recursion & 2 & 1 & 1 & 1 & 0 & 2 \\
CWE-754 & Improper Check for Unusual or Exceptional Conditions & 2 & 0 & 2 & 0 & 2 & 1 \\
CWE-770 & Allocation of Resources Without Limits or Throttling & 2 & 0 & 2 & 2 & 0 & 2 \\
CWE-824 & Access of Uninitialized Pointer & 2 & 0 & 2 & 1 & 1 & 2 \\
CWE-863 & Incorrect Authorization & 2 & 0 & 2 & 0 & 2 & 2 \\
CWE-126 & Buffer Over-read & 1 & 0 & 1 & 1 & 0 & 1 \\
CWE-203 & Observable Discrepancy & 1 & 0 & 1 & 0 & 1 & 1 \\
CWE-209 & Information Exposure Through an Error Message & 1 & 0 & 1 & 1 & 0 & 1 \\
CWE-227 & Improper Fulfillment of API Contract ('API Abuse') & 1 & 0 & 1 & 1 & 0 & 1 \\
CWE-271 & Privilege Dropping / Lowering Errors & 1 & 0 & 1 & 1 & 0 & 1 \\
CWE-354 & Improper Validation of Integrity Check Value & 1 & 0 & 1 & 0 & 1 & 1 \\
CWE-475 & Undefined Behavior for Input to API & 1 & 0 & 1 & 1 & 0 & 1 \\
CWE-485 & Insufficient Encapsulation & 1 & 0 & 1 & 1 & 0 & 1 \\
CWE-532 & Insertion of Sensitive Information into Log File & 1 & 0 & 1 & 0 & 1 & 1 \\
CWE-665 & Improper Initialization & 1 & 0 & 1 & 1 & 0 & 1 \\
CWE-670 & Always-Incorrect Control Flow Implementation & 1 & 0 & 1 & 0 & 1 & 1 \\
CWE-680 & Integer Overflow to Buffer Overflow & 1 & 0 & 1 & 1 & 0 & 1 \\
CWE-682 & Incorrect Calculation & 1 & 0 & 1 & 0 & 1 & 1 \\
CWE-74 & Improper Neutralization of Special Elements in Output Used by a Downstream Component ('Injection') & 1 & 0 & 1 & 1 & 0 & 1 \\
CWE-762 & Mismatched Memory Management Routines & 1 & 0 & 1 & 1 & 0 & 1 \\
CWE-77 & Improper Neutralization of Special Elements used in a Command  & 1 & 0 & 1 & 1 & 0 & 1 \\
CWE-823 & Use of Out-of-range Pointer Offset & 1 & 0 & 1 & 1 & 0 & 1 \\
CWE-95 & Improper Neutralization of Directives in Dynamically Evaluated Code ('Eval Injection') & 1 & 0 & 1 & 1 & 0 & 1 \\
\bottomrule
\end{tabular}
\label{tab:app-cwes}
\end{table*}

\subsection{Generated Instance Characteristics}

Table~\ref{tab:app-attributes} reports the size of the artifacts that make up a weakness instance. Injected edits are small and local, which helps them stay latent: 1\,025 of 1\,034 instances changed a single file, 944 confined the change to one hunk, that is one contiguous block of changed lines with its surrounding context.

Every instance ships a PoV script to trigger the vulnerability, which implementation differs depending on the pipeline (Table~\ref{tab:app-pov}). The fuzzer-guided pipeline replays an input recovered from a seeded libFuzzer session. The agentic pipeline ships a fixed input file the agent constructed, along with a driving program.

\begin{table}[H]
\centering
\caption{Size of the artifacts that make up an instance. Micro-averages over all instances, not grouped by project; the final row is per project.}
\small
\begin{tabular}{@{}lrrrr@{}}
\toprule
 & \textbf{Med.} & \textbf{Mean} & \textbf{p90} & \textbf{Max} \\
\midrule
\multicolumn{5}{@{}l}{\emph{Injected diff}} \\
\quad Lines added & 1 & 1.19 & 2 & 17 \\
\quad Lines deleted & 1 & 2.79 & 4 & 643 \\
\quad Lines changed & 2 & 3.98 & 5 & 644 \\
\quad Hunks & 1 & 1.52 & 1 & 337 \\
\quad Files & 1 & 1.01 & 1 & 3 \\
\midrule
\multicolumn{5}{@{}l}{\emph{Verification artifacts}} \\
\quad PoV entry script (lines) & 20 & 22.69 & 49 & 185 \\
\quad Project test suite (tests) & 41 & 281.60 & 633 & 6\,024 \\
\bottomrule
\end{tabular}
\label{tab:app-attributes}
\end{table}

\begin{table}[H]
\centering
\caption{Form of the released proof of vulnerability. \textbf{Lines}: mean length of the \texttt{exploit.sh} entry point.}
\small
\begin{tabular}{@{}lrrrr@{}}
\toprule
\textbf{PoV form} & \textbf{All} & \textbf{P1} & \textbf{P2} & \textbf{Lines} \\
\midrule
fuzz-discovered replay & 421 & 421 & 0 & 4 \\
script-constructed input & 341 & 222 & 119 & 36 \\
fixed input file & 272 & 0 & 272 & 35 \\
\bottomrule
\end{tabular}
\label{tab:app-pov}
\end{table}

\subsection{Resemblance to Real Vulnerabilities}

The main paper reports a Kolmogorov--Smirnov distance of 0.165 between the injected vulnerabilities and real ones, compared with a real-to-real baseline of 0.190. This section describes how those values were obtained.

For an edit statistic, let $x_1,\dots,x_n$ denote the observations from the injected corpus and $y_1,\dots,y_m$ those from a reference corpus of real vulnerabilities. If $F_n$ and $G_m$ are their empirical distribution functions, the two-sample Kolmogorov--Smirnov statistic is
\[ D_{n,m} \;=\; \sup_{x}\ \bigl|F_n(x) - G_m(x)\bigr| , \]
It measures the largest gap between the two empirical distributions. A value of $0$ means that the observed distributions coincide, while larger values indicate greater separation. The statistic requires no parametric model, which is useful because patch-size distributions are strongly skewed.

The reference corpus contains all 300 SEC-bench real-world instances: 200 from the \texttt{cve} split and 100 from the \texttt{oss} split. We compute the statistic along five axes: functions changed, files changed, hunks, changed lines, and added lines. The first three capture the structure of an edit; the last two capture its size.

A KS distance has no universal threshold for similarity. We therefore compare the SEC-bench \texttt{oss} and \texttt{cve} splits against each other. Since both contain real vulnerability patches, but still come from two separate data sources, their distance provides a practical baseline for the variation between two real corpora. An injected-to-real distance near or below this baseline is therefore not, by itself, evidence that the injected edits are distinguishable as artificial.

\begin{figure*}[h]
\centering
\includegraphics[width=0.9\textwidth]{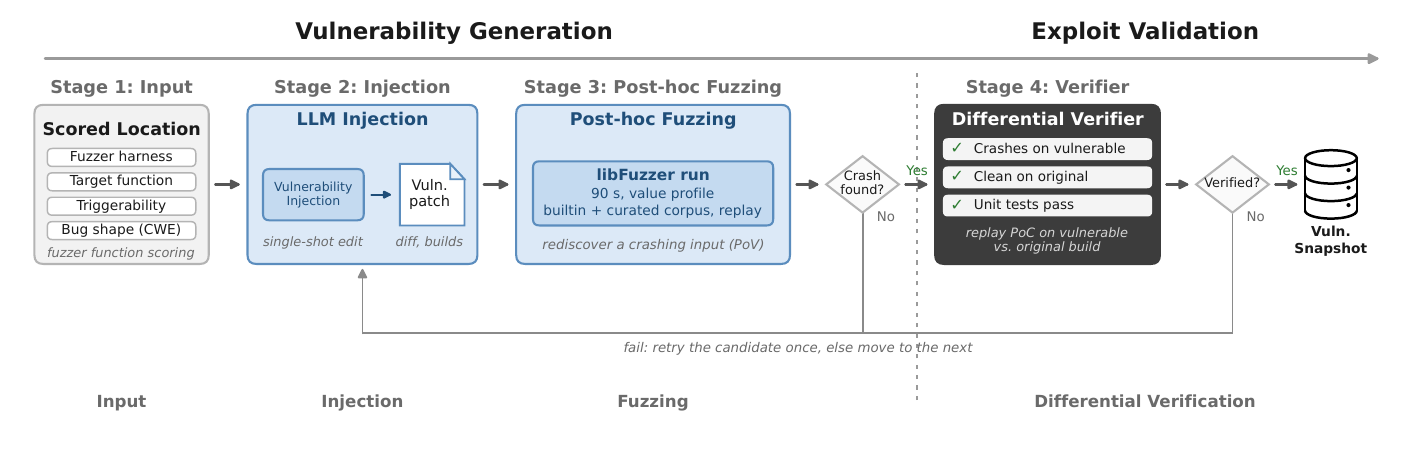}
\caption{\textbf{P1}: CyberForge's fuzzer-guided injection pipeline.}
\label{fig:app-p1-flow}
\end{figure*}

\begin{figure*}[h]
\centering
\includegraphics[width=0.9\textwidth]{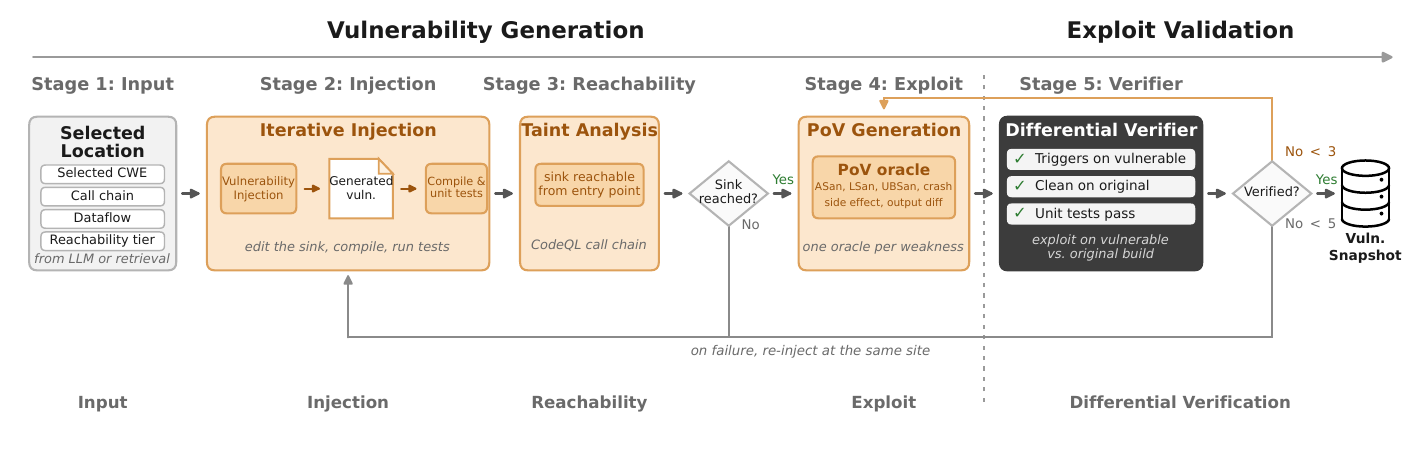}
\caption{\textbf{P2}: CyberForge's agentic in-context injection pipeline.}
\label{fig:app-p2-flow}
\end{figure*}

For functions modified, the injected corpus has a distance of 0.165, below the real-to-real baseline of 0.190. The corresponding distances for files touched and hunk count are 0.123 and 0.243, compared with baselines of 0.065 and 0.205. Thus, the injected edits broadly resemble real patches in where and how they are distributed across the codebase.

\section{Injection Pipelines}
\label{app:pipelines}

Both pipelines start from a selected candidate location and terminate at the same differential oracle.

The fuzzer-guided pipeline (Figure~\ref{fig:app-p1-flow}) considers only functions reachable from an existing libFuzzer harness and ranks them by triggerability. The model performs a single-shot edit, after which the harness fuzzes the modified project for 90\,s using a seeded corpus. A failed candidate is retried once before the pipeline moves on.

The agentic pipeline (Figure~\ref{fig:app-p2-flow}) iteratively injects a weakness at a selected location, using compiler and unit-test feedback. Once the project builds and its tests pass, a taint-analysis agent examines CodeQL-extracted dataflow from the source (an entry point through which external input enters the program) to the sink (a code location where that input is used in a potentially unsafe operation). The agent autonomously explores the dataflow graph backward from the sink, iteratively identifies inputs that could trigger the vulnerable behavior, and verifies that an external-input source can still reach the injected sink. If no such path exists, the analysis context is returned to the injection agent as feedback. Otherwise, a proof-of-vulnerability (PoV) agent constructs a trigger using strategies tailored to the weakness type.

Both pipelines use Gemma 4 31B through mini-swe-agent. No sampling parameters are overridden, so generation uses the model defaults: temperature 1.0, top-$p$ 0.95, and top-$k$ 64. Each agent invocation is capped at 200 iterations.

\section{Generation Cost}
\label{app:cost}

This section reports the cost of generating the CyberForge corpus and collecting the teacher trajectories.

We account for self-hosted inference in two ways: the electricity consumed during generation and the estimated cost of processing the same number of tokens through a hosted API. For API-based inference, we report the charges recorded by the provider.

The accounting assumes $2\times$H100 SXM at 700\,W, CPU nodes at 400\,W, a PUE of 1.3, and average electricity cost at \$0.1359/kWh. Hosted inference for a Gemma-class model is priced at OpenRouter's published rate of \$0.09 per million input tokens and \$0.34 per million output tokens (2026-07-31). During our experiment GPT-5.4-mini was billed at \$0.0375 per million cached input tokens and \$2.25 per million output tokens.

\begin{figure*}[p]
\centering
\includegraphics[width=0.45\textwidth]{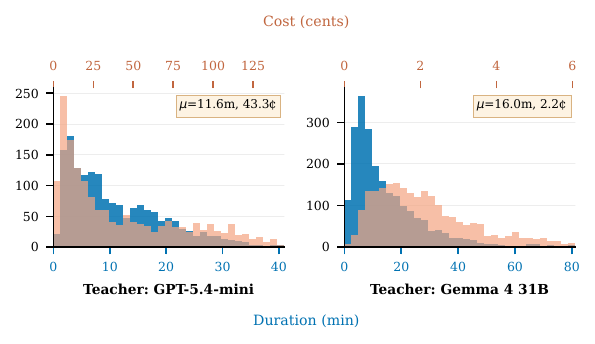}
\caption{Time and cost per teacher trajectory generation run. Duration is read on the lower
axis, cost on the upper one. Boxes give the means; axes are clipped at the 99th percentile.}
\label{fig:app-time-cost}
\end{figure*}

\begin{figure*}[p]
\centering
\includegraphics[width=0.72\textwidth]{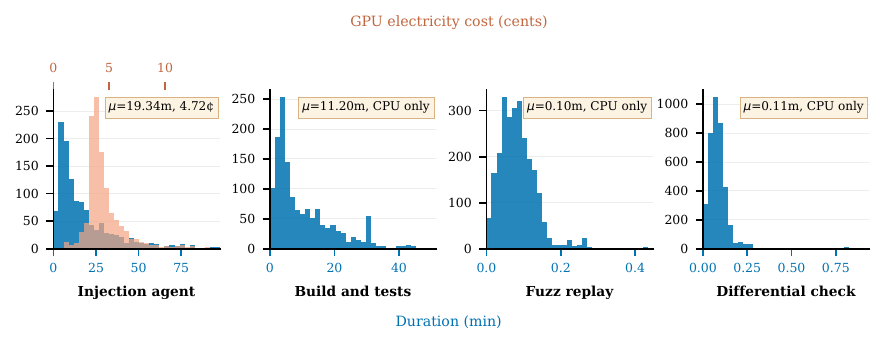}
\caption{Time and cost breakdown of the P1 fuzzer-guided injection pipeline per attempt. The injection agent is the only stage that calls the model; the rest run on CPU.}
\label{fig:app-p1-stages}
\end{figure*}

\begin{figure*}[p]
\centering
\includegraphics[width=0.75\textwidth]{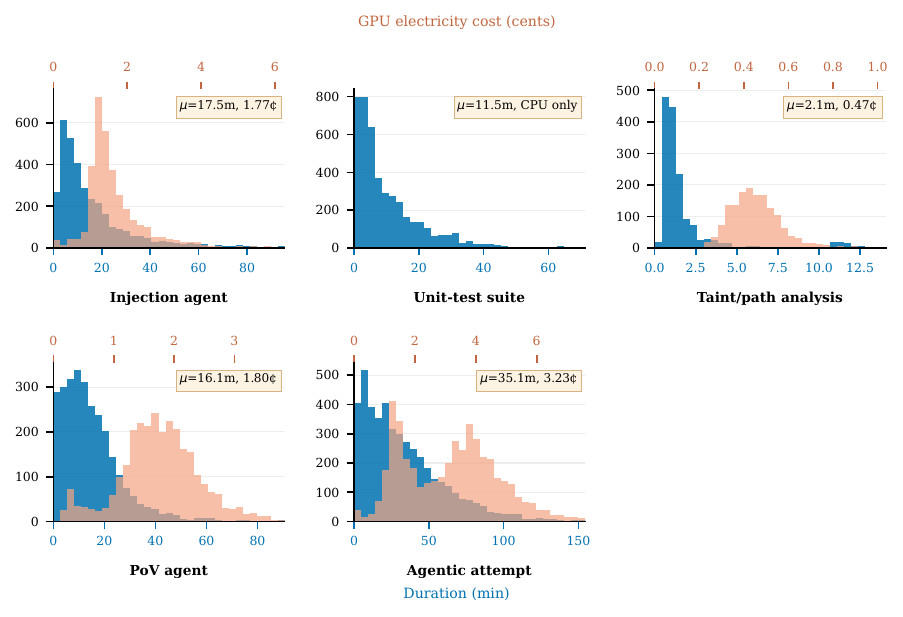}
\caption{Time and cost per sub-agent of the P2 agentic injection pipeline per attempt.}
\label{fig:app-subagents}
\end{figure*}

GPU time covers model inference. CPU time covers the per-project Docker containers used to build each target, run the fuzzing harnesses, and execute the unit tests. Container execution incurs no token cost, but its energy use is included in the metered electricity estimate. By contrast, the API-equivalent estimate prices model tokens only and excludes all container work, so it does not capture the full cost of running the pipeline on hosted infrastructure.

Table~\ref{tab:app-compute} reports the GPU and CPU time used for vulnerability injection and teacher-trajectory generation. Table~\ref{tab:app-tokens} reports token usage at each stage of the two pipelines.

\begin{table}[h]
\centering
\caption{Compute used for corpus generation and teacher-trajectory collection. CPU hours are machine-hours over the union of active intervals, so concurrent runs count once. The $2\times$H100 endpoint was shared across activities, so GPU wall time is not exclusive to one row.}
\scriptsize
\setlength{\tabcolsep}{3.6pt}
\begin{tabular}{@{}lrrrrr@{}}
\toprule
 & \multicolumn{2}{c}{\textbf{Hours}} & \multicolumn{2}{c}{\textbf{Electricity}} & \\
\cmidrule(lr){2-3}\cmidrule(lr){4-5}
\textbf{Activity} & \textbf{CPU} & \textbf{GPU} & \textbf{CPU} & \textbf{GPU} & \textbf{Parallelism} \\
\midrule
Injection             & 874 & 775 & \$62 & \$192 & 4.7--13.0$\times$ \\
Teacher: Gemma 4 31B  & 97  & 97  & \$7  & \$24  & 6.3$\times$ \\
Teacher: GPT-5.4-mini & 129 & --- & \$9  & ---   & 2.7$\times$ \\
\bottomrule
\end{tabular}
\label{tab:app-compute}
\end{table}

\begin{table}[h]
\centering
\caption{Token usage for each type of task and phase.}
\small
\begin{tabular}{@{}lrr@{}}
\toprule
\textbf{Activity / phase} & \textbf{Input} & \textbf{Output} \\
\midrule
\multicolumn{3}{@{}l}{\emph{Corpus generation}} \\
\quad P1 injection agent  & 958.8\,M & 15.47\,M \\
\quad P1 fuzz replay      & \multicolumn{2}{c}{\emph{no model stage}} \\
\quad P2 injection agent  & 1\,668.2\,M & 30.84\,M \\
\quad P2 PoV generation   & 819.3\,M     & 19.55\,M \\
\quad P2 taint analysis   & 286.7\,M     & 5.58\,M \\
\quad P2 other agents     & 116.1\,M     & 2.00\,M \\
\midrule
\multicolumn{3}{@{}l}{\emph{Teacher trajectory collection}} \\
\quad GPT-5.4-mini        & 2\,871.7\,M & 143.23\,M \\
\quad Gemma 4 31B         & 526.1\,M     & 4.07\,M \\
\midrule
\textbf{Total}            & \textbf{7\,246.8\,M} & \textbf{220.74\,M} \\
\bottomrule
\end{tabular}
\label{tab:app-tokens}
\end{table}

Table~\ref{tab:app-money} summarizes the estimated cost of CyberForge. Corpus generation consumed \$253 in electricity; pricing the same token volume at hosted-inference rates, while retaining the cost of running the containers, raises the estimate to \$433. Collecting the teacher trajectories cost \$748, including \$738.62 in measured API charges for GPT-5.4-mini and Gemma teacher runs consumed \$31 in electricity. The total cost of generating the dataset and collecting teacher trajectories was \$1\,032, compared with an estimated \$1\,237 if all inference tokens had been purchased through hosted services.
Amortized over the validated corpus, generation cost \$0.25 per instance in electricity, or \$0.42 when inference is valued at hosted rates. The corresponding fully API-billed costs are \$0.87 per instance for SEC-bench and \$2.77 for CVE-Genie.

Figure~\ref{fig:app-time-cost} gives the distributions for teacher trajectories, which take 9 to 11 minutes at the median. Figure~\ref{fig:app-p1-stages} breaks the fuzzer-guided attempt down by stage, where the injection agent and the build-and-test gate dominate and the replay and differential checks take about five seconds each, and Figure~\ref{fig:app-subagents} the agentic attempt by sub-agent, with a median of 27 minutes. Both exclude the 9.6\,\% of model calls that returned an error before producing any output.

\begin{table}[h]
\centering
\caption{CyberForge cost estimation for each task type including CPU electricity cost. \textbf{Actual}: paid electricity price (or token-cost for GPT-5.4-mini). \textbf{API}: token-based cost (estimated for Corpus generation and Gemma teacher). \textbf{Per instance} divides by the 1\,034 validated instances.}
\scriptsize
\setlength{\tabcolsep}{4pt}
\begin{tabular}{@{}lrrrr@{}}
\toprule
 & \multicolumn{2}{c}{\textbf{Total}} & \multicolumn{2}{c}{\textbf{Per instance}} \\
\cmidrule(lr){2-3}\cmidrule(lr){4-5}
\textbf{Activity} & \textbf{Actual} & \textbf{API} & \textbf{Actual} & \textbf{As API} \\
\midrule
Corpus generation        & \$253     & \$433     & \textbf{\$0.25} & \$0.42 \\
Teacher: GPT-5.4-mini    & \$748     & \$748     & \textbf{\$0.72} & \$0.72 \\
Teacher: Gemma 4 31B     & \$31      & \$56      & \textbf{\$0.03} & \$0.05 \\
\midrule
\textbf{Total}           & \$1\,032 & \$1\,237 & \$1.0 & \$1.2 \\
\bottomrule
\end{tabular}
\label{tab:app-money}
\end{table}

\section{Supervised Fine-Tuning Details}
\label{app:sft}

We fine-tune Gemma~4 at three scales (E4B, 12B and 31B) on the teacher trajectories collected over the corpus, training one student per base model and teacher. All students use the same LoRA recipe: the base weights stay frozen and low-rank adapters are attached to the attention and MLP projections of the language model only, leaving the vision tower untouched. Table~\ref{tab:app-sft} lists the hyperparameters.

\begin{table}[h]
\centering
\caption{Fine-tuning configuration, identical across all six student/teacher combinations.}
\small
\begin{tabular}{@{}ll@{}}
\toprule
\textbf{Hyperparameter} & \textbf{Value} \\
\midrule
Base models        & Gemma~4 E4B / 12B / 31B \\
Adaptation         & LoRA, base frozen \\
LoRA rank $r$      & 32 \\
LoRA $\alpha$      & 64 \\
LoRA dropout       & 0.05 \\
Adapted modules    & attention $+$ MLP projections \\
Precision          & BFloat16 \\
Attention          & SDPA \\
Learning rate      & 1e-4 \\
Weight decay       & 0.0 \\
Schedule           & Cosine \\
Warmup ratio       & 0.03 \\
Optimizer          & AdamW (fused) \\
Epochs             & 3 \\
Batch size         & 1 \\
Gradient accumulation & 2 \\
Gradient clipping  & 1.0 \\
Gradient checkpointing & Enabled \\
Hardware           & 1$\times$H200 (147\,GB) or H100 \\
Seed               & 42 \\
\bottomrule
\end{tabular}
\label{tab:app-sft}
\end{table}

\paragraph{Teacher-trajectory cleansing.} Raw teacher trajectories are recorded in the teacher's own generation environment and are not directly usable for training a SEC-bench student, so we pass them through a cleansing pipeline before fine-tuning. (i) We retain only oracle-verified successful trajectories. (ii) We align the teacher environment to the SEC-bench evaluation harness \cite{secbench}: repro/build invocations are normalized to secb repro/secb build, exploit-artifact paths are remapped to /testcase, and ripgrep calls are rewritten to grep (the evaluation container lacks the ripgrep package). (iii) Each trajectory is linearized to one command per assistant turn. (iv)  The teacher's reasoning is retained, after removing its boilerplate bold-header summary. (v) Tool outputs are re-rendered through the evaluation observation template (a 10\,k-character head/tail cap) and long observation histories are compressed with a sliding window (recent observations kept in full, older ones reduced to short stubs) so each trajectory fits the training context without ever truncating the assistant turn that writes the patch. 

\noindent
In the main paper we show that our fine-tuned models maintain some gains even in out-of-distribution setting with different evaluation environment (keeping the mini-swe-agent scaffold)

\section{Example of Injected Vulnerability}
\label{app:example}

This section presents one instance of the generated corpus, \texttt{guetzli/\allowbreak vulnerability\_\allowbreak FZ\_\allowbreak 24}: the weakness CyberForge introduced, the input that triggers it and explanations about this vulnerability.

Guetzli is Google's JPEG compressor, used by image hosts, CDNs, and web developers to reduce image file sizes before delivery.

\subsection{The Vulnerability}

A JPEG file consists of a series of segments containing either compressed image data or metadata (e.g. color profiles and camera settings). Each segment begins with a two-byte length field that tells the decoder how many bytes belong to it.

In \texttt{ProcessAPP} (Listing~\ref{lst:diff}), the original code validates this field in two steps. \texttt{VERIFY\_INPUT} checks that the declared length is valid under the JPEG format, between 2 and 65535. \texttt{VERIFY\_LEN} then checks that the input buffer actually contains that many bytes. CyberForge removed the buffer-length check.

As a result, a JPEG may declare a segment length of up to 65535 bytes while providing much less data. The \texttt{std::string} constructor uses the declared length when copying from the segment, causing it to read past the end of the input buffer. This is an out-of-bounds read, CWE-125, because the attacker controls the copy length without having to supply the corresponding number of bytes.

\subsection{Triggering the Vulnerability}

The trigger is a 504-byte file that opens with the bytes:

\begin{center}\small\ttfamily
FF D8 FF E0 FF FF 4A 46 49 46
\end{center}

\noindent
The file starts normally: \texttt{FF D8} is the JPEG start marker, and \texttt{FF E0} begins an APP0 metadata segment. The next two bytes, \texttt{FF FF}, declare a segment length of 65\,535 bytes. This is the largest value accepted by \texttt{VERIFY\_INPUT}.
Only 504 bytes are present, so the removed \texttt{VERIFY\_LEN} check would have rejected the file. Without it, the code proceeds using the declared length. The \texttt{std::string} constructor then attempts to copy 65\,536 bytes from an offset three bytes into the buffer. Since only 501 bytes remain, the read extends 65\,035 bytes beyond the end of the file.

\begin{listing}[h]
\begin{lstlisting}[numbers=none,moredelim={[il][\color{red}]{@}}]
bool ProcessAPP(const uint8_t* data,
                const size_t len,
                size_t* pos, JPEGData* jpg) {
  VERIFY_LEN(2);
  size_t marker_len = ReadUint16(data, pos);
  VERIFY_INPUT(marker_len, 2, 65535,
               MARKER_LEN);
@-  VERIFY_LEN(marker_len - 2);
  // Save the marker type with the app data.
  std::string app_str(
      reinterpret_cast<const char*>(
          &data[*pos - 3]), marker_len + 1);
  *pos += marker_len - 2;
  jpg->app_data.push_back(app_str);
  return true;
}
\end{lstlisting}
\caption{Example of vulnerability injected by CyberForge in the Guetzli repository. The line in red is the one the injection deletes; nothing else changes.}
\label{lst:diff}
\end{listing}

\subsection{Why the Defect Is Realistic}

This bug follows the same basic pattern as Heartbleed (CVE-2014-0160): an attacker supplies a length larger than the accompanying data, and the program reads beyond that data into adjacent memory. File formats with embedded length fields are especially prone to this class of error, including image metadata segments.

The defect is also easy to miss. The attacker-controlled value is still checked against the JPEG format's permitted range shortly before it is used, which can make the code appear adequately validated during review. The missing check concerns whether the declared length fits within the actual input buffer.

Ordinary unit tests are unlikely to catch the problem. JPEGs produced by conforming encoders contain segment lengths that match the available data, so the removed check has no visible effect on valid files. Triggering the bug requires a deliberately malformed JPEG with a valid length value but insufficient segment data.

\subsection{Impact}

The out-of-bounds read could disclose server memory. In an unsanitized production build, the copied bytes become part of \texttt{app\_data} and are written into the re-encoded JPEG returned to the uploader. Because the segment length is 16 bits, a single request can expose nearly 64,KB beyond the input buffer.

The leaked region may contain data from other work handled by the same process, including image buffers or metadata belonging to other users. Depending on the heap layout, it could also contain session tokens, API keys, or other secrets. Repeated uploads may reveal different regions as allocations change.

The request need not look like a failure. The service can accept the upload, return a valid JPEG, and record the operation as successful while unintentionally including process memory in the response.

\subsection{Directory Structure and File Contents}
Every instance ships as a self-contained directory with the following layout. The files that define this instance are reproduced below.

\begin{figure*}[h]
\centering
\begin{corpustree}
guetzli/
|-- project.json                     # language, upstream URL, secure base commit
|-- setup/                           # OSS-Fuzz build
|   |-- Dockerfile
|   |-- build.sh
|   `-- project.yaml
|-- unit_tests/                      # non-regression gate
|   |-- test.sh                      # standardized entrypoint
|   `-- parse_results.py             # -> {"passed": N, "failed": M}
`-- vulnerabilities/
    |-- vulnerability_FZ_0/          # one directory per instance
    |   ...
    `-- vulnerability_FZ_24/         # the instance detailed below
        |-- inject_vulnerability.diff     # apply -> introduces the vulnerability
        |-- vulnerability_metadata.json   # id, cwe_id, secure base commit
        |-- sanitizer_report.txt          # sanitizer crash proving it triggers
        `-- exploit_files/
            |-- exploit.sh                # proof-of-vulnerability entrypoint
            |-- fuzz_poc.py               # PoV driver
            `-- generated_inputs/         # seed corpus
\end{corpustree}
\end{figure*}

\begin{figure*}[h]
\centering
\begin{corpusfile}{project.json}
{
  "project": "guetzli",
  "main_repo_url": "https://github.com/google/guetzli",
  "target_dir": "guetzli",
  "secure_base_commit": "214f2bb42abf5a577c079d00add5d6cc470620d3",
  "unit_tests": {"enabled": true, "expected_passing_count": 10},
  "language": "c++"
}
\end{corpusfile}
\end{figure*}

\begin{figure*}[h]
\centering
\begin{corpusfile}{inject\_vulnerability.diff}
diff --git a/guetzli/jpeg_data_reader.cc b/guetzli/jpeg_data_reader.cc
@@ -398,7 +398,7 @@ bool ProcessAPP(const uint8_t* data, size_t* pos, ...)
   VERIFY_LEN(2);
   size_t marker_len = ReadUint16(data, pos);
   VERIFY_INPUT(marker_len, 2, 65535, MARKER_LEN);
-  VERIFY_LEN(marker_len - 2);
+
   // Save the marker type together with the app data.
   std::string app_str(reinterpret_cast<const char*>(
       &data[*pos - 3]), marker_len + 1);
\end{corpusfile}
\end{figure*}

\begin{figure*}[h]
\centering
\begin{corpusfile}{vulnerability\_metadata.json}
{
  "id": "vulnerability_FZ_24",
  "project": "guetzli",
  "producer": "fuzz_poc_guided",
  "cwe_id": "CWE-125",
  "cwe_group": "Post buffer operation",
  "secure_base_commit": "214f2bb42abf5a577c079d00add5d6cc470620d3"
}
\end{corpusfile}
\end{figure*}

\begin{figure*}[h]
\centering
\begin{corpusfile}{exploit\_files/exploit.sh}
#!/bin/bash
set -euo pipefail
cd /src/guetzli
python3 exploit_files/fuzz_poc.py
\end{corpusfile}
\end{figure*}

\begin{figure*}[h]
\centering
\begin{corpusfile}{exploit\_files/fuzz\_poc.py (excerpt)}
fuzzer = CFG["selected_fuzzer"]["fuzzer"]
binary = Path("/out") / fuzzer              # the built OSS-Fuzz target
...
def run_and_capture(args, log_name):
    proc = subprocess.run(args, stdout=subprocess.PIPE,
                          stderr=subprocess.STDOUT, text=True)
    output = proc.stdout
    crash = ("AddressSanitizer" in output
             or "UndefinedBehaviorSanitizer" in output
             or "runtime error:" in output)
    return proc, crash, ...
...
# time-boxed libFuzzer campaign over the seed corpus
args = [str(binary),
        f"-max_total_time={CFG.get('max_total_time', 60)}",
        f"-max_len={max_len}", f"-dict={dict_path}", str(CORPUS_DIR)]
proc = subprocess.run(args, stdout=subprocess.PIPE,
                      stderr=subprocess.STDOUT, text=True)
...
asan_like = ("AddressSanitizer" in proc.stdout
             or "runtime error:" in proc.stdout)
crash = bool(crash_artifacts) or (asan_like and not oom and not leak)
print("TRIGGERED" if crash else "NOT_TRIGGERED")
\end{corpusfile}
\end{figure*}

\begin{figure*}[h]
\centering
\begin{corpusfile}{exploit\_files/generated\_inputs/ -- triggering input (504 bytes, hexdump)}
00000000  ff d8 ff e0 ff ff 4a 46  49 46 00 01 01 02 00 1c   ......JFIF......
00000010  00 1c 00 00 ff db 00 43  00 28 1c 1e 23 1e 19 28   .......C.(..#..(
00000020  23 21 23 2d 2b 28 30 3c  64 41 3c 37 37 3c 7b 58   #!#-+(0<dA<77<{X
...
# 504 bytes total. Bytes 0-1 = SOI (ff d8); bytes 2-3 begin an APP0
# segment whose length field (bytes 4-5) is ff ff = 65535, far more
# than the 504 bytes present -> ProcessAPP reads past the input buffer.
\end{corpusfile}
\end{figure*}

\begin{figure*}[h]
\centering
\begin{corpusfile}{sanitizer\_report.txt (excerpt)}
==432==ERROR: AddressSanitizer: heap-buffer-overflow
READ of size 65531 at 0x6fc93620b1f8 thread T0
    #0 ProcessAPP jpeg_data_reader.cc:403:15
SUMMARY: AddressSanitizer: heap-buffer-overflow
    jpeg_data_reader.cc:403:15 in guetzli::ProcessAPP
\end{corpusfile}
\end{figure*}

\end{document}